%% file: CGI_v3.tex
\documentclass[11pt,a4paper]{article}

\usepackage[utf8]{inputenc}
\usepackage[english]{babel}
\usepackage{latexsym,mathtools, amsfonts, amssymb}
\usepackage{cite}
\usepackage{xspace}
\usepackage{bbm}
\usepackage{graphicx}
\usepackage{xcolor} 

\usepackage{fullpage}
\usepackage[font=small]{caption}

\numberwithin{equation}{section}

\newenvironment{formula}%
{\begin{equation}\begin{aligned}\relax}%
{\end{aligned}\end{equation}\ignorespacesafterend}

\usepackage{tensind}
\tensordelimiter{;}
\tensorformat{s}

\begingroup\lccode`~=``
\lowercase{\endgroup\def~}#1 {;#1;}
\mathcode``="8000
\makeatletter
\g@addto@macro\bfseries{\boldmath}
\makeatother

\usepackage{hyperref}

\input{definitions}

\begin{document}

\begin{titlepage}
\begin{flushright} 
\small
Nikhef-2016-014
\end{flushright}
\vspace*{1cm}

\begin{center}
    {\LARGE\bfseries The exceptional story of massive IIA supergravity}
\\[10mm]
\textbf{Franz Ciceri,\, Adolfo Guarino} \,and\, \textbf{Gianluca Inverso}\\[5mm] 
\vskip 4mm
{\em Nikhef Theory Group, Science Park 105, 1098 XG Amsterdam, The
  Netherlands}
\end{center}

\vspace{15ex}

\begin{center}
{\bfseries Abstract}
\end{center}
\begin{quotation} 
\noindent The framework of exceptional field theory is extended by introducing consistent deformations of its generalised Lie derivative. For the first time, massive type~IIA supergravity is reproduced geometrically as a solution of the section constraint. 
This provides a unified description of all ten- and eleven-dimensional maximal supergravities. 
The action of the $\,\Eseven$ deformed theory is constructed, and reduces to those of exceptional field theory and gauged maximal supergravity in respective limits.
The relation of this new framework to other approaches for generating the Romans mass non-geometrically is discussed.
\end{quotation}

\vfill
\blfootnote{Emails: \texttt{f.ciceri@nikhef.nl}\,,\ \texttt{a.guarino@nikhef.nl}\,,\ \texttt{g.inverso@nikhef.nl}\,.}
\end{titlepage}


\tableofcontents

\setcounter{footnote}{0}

\section{Motivation and outlook}
\label{sec:Intro}

Exceptional field theory (EFT) provides a unified framework where to describe massless type~II and eleven-dimensional supergravity \cite{Hohm:2013vpa,Hohm:2013uia,Hohm:2014fxa,Hohm:2015xna,Abzalov:2015ega,Musaev:2015ces,Berman:2015rcc}.
It is therefore natural to ask whether the unifying abilities of EFT could also allow for an implementation of the massive IIA theory \cite{Romans:1985tz}. 
In EFT, $\,\En$ covariance is made manifest by adding extra internal coordinates to the ten- or eleven-dimensional spacetime in order to gain new insights into the structure of string/M-theory. 
Consistency of the theory eventually requires to impose a \textit{section constraint} which restricts all fields to depend at most on ten or eleven physical coordinates.
After solving the section constraint, EFT reduces (locally) to an exceptional generalised geometry (EGG) formulation of massless type II or eleven-dimensional supergravities \cite{Coimbra:2011ky,Coimbra:2012af}.
Applications range from the study of consistent truncations \cite{Hohm:2014qga,Baguet:2015sma,Malek:2015hma} to loop computations of higher derivative corrections to the M-theory effective action \cite{Bossard:2015foa}.

While the embedding of the eleven-dimensional and massless type II supergravities into EFT is well understood, the one of massive IIA supergravity remains so far an open question.
In fact, a puzzle arises when facing this issue.
On the one hand, being a fully consistent ten-dimensional maximal supergravity in its own right, massive IIA should posses an associated EGG capturing its degrees of freedom and local symmetries in the same fashion as for the massless type II theories.
It is therefore natural to expect that such a generalised geometry would descend from EFT after choosing some specific solution of the section constraint.
On the other hand, solutions of the section constraint in EFT have been classified and are known to exclusively correspond to the massless type II and eleven-dimensional supergravities \cite{Berman:2010is,Berman:2012vc,Blair:2013gqa,Bossard:2015foa,Bandos:2015rvs}.
It thus seems that some violation of the section constraint is needed in order to reproduce the Romans mass. 
In the context of double field theory (DFT), the Romans mass was implemented by
allowing a Ramond--Ramond (RR) potential to depend on a non-geometric (winding) coordinate \cite{Hohm:2011cp}, thus again suggesting that a similar scenario should take place in EFT.
In this case, however, there would be no direct relation with an EGG for massive IIA in ten dimensions.
This paper provides a solution to this puzzle and, in doing so, unveils an extension of the EFT framework.

The Romans mass $\,m_{\rm R}\,$ has always manifested itself as a deformation parameter in any construction related to type IIA. This is the case, for instance, for the supersymmetric AdS vacua of \cite{Behrndt:2004km,Behrndt:2004mj,Lust:2004ig}. When considering dual holographic models, the Romans mass translates into a deformation of the field theory in the form of a Chern-Simons term with level $\,k\,$ given by $\,k/(2 \pi l_{s})=m_{\rm R}\,$ \cite{Schwarz:2004yj,Guarino:2015jca} (see also \cite{Gaiotto:2009mv,Gaiotto:2009yz}). A more recent example involves the consistent reduction of the massive IIA theory on the six-sphere \cite{Guarino:2015jca,Guarino:2015vca}. In this case it was shown \cite{Guarino:2015jca,Guarino:2015qaa} that, after truncation to four dimensions, the Romans mass appears as an electric-magnetic deformation parameter of the types constructed in \cite{DallAgata:2011aa,Dall'Agata:2012bb} and classified in \cite{Dall'Agata:2014ita,Inverso:2015viq}. 
These results suggest that, in order to embed the massive IIA theory in EFT, one should investigate the possible deformations of the latter.

In this paper, we will show that EFT does admit consistent deformations which still allow for ten- and/or eleven-dimensional solutions of the section constraint.
For one of these deformations, there exists a purely geometric ten-dimensional solution which precisely corresponds to massive IIA supergravity, and thus defines as a byproduct the associated EGG.
This new deformed EFT framework endows massive IIA supergravity with the same geometrical and group-theoretical tools so far exclusive to the massless theories.

We now present a brief summary of the structure of the deformed EFT framework.
EFT is based on an `external' spacetime and an `internal' extended space with coordinates $\,x^\mu\,$ and $\,y^{M}\,$, where $\,\mu=0,\ldots,D-1\,$, $\,{M=1,\ldots,\textrm{dim }\mathcal{R}_{\rm v}}\,$ and $\,\mathcal{R}_{\rm v}\,$ denotes the $\,\En\,$ representation of the vector fields in the theory (see Table~\ref{Table:intro_irreps}). 
Internal generalised diffeomorphisms act on fields by means of a generalised Lie derivative $\,\mathbb{L}_{\Lambda}\,$.
While all fields and parameters formally depend on the full set of coordinates $\,(x^\mu,\,y^M)\,$, the dependence on the internal coordinates is ultimately restricted to a physical subset by the section constraint
\begin{equation}
\label{intro:Y-cond}
Y^{PQ}{}_{MN} \, \partial_{P} \, \otimes \, \partial_{Q} = 0 \ ,
\end{equation}
where $\,\partial_M\equiv\frac{\partial}{\partial y^M}\,$ and $\,`Y^PQ_MN \,$ is a specific $\,\En\times\bbR^+\,$ invariant tensor \cite{Berman:2012vc}.
After choosing a maximal solution of this constraint, EFT effectively reduces to eleven-dimensional or type IIB supergravity in a $\,{D+n}\,$ or $\,D+(n-1)\,$ dimensional split, respectively. Such a split of the physical coordinates into the $D$-dimensional external spacetime and the $n$- or $(n-1)$-dimensional internal space explicitly breaks the Lorentz covariance of the eleven- or ten-dimensional theory but does not truncate any of its degrees of freedom. 
The generalised Lie derivative then encodes the ordinary internal diffeomorphisms and $p$-form gauge transformations of the physical theory in the corresponding dimensional split.

The central result of this work is the construction of `$X$ deformed' exceptional field theories (XFT's) based on the following modification of the generalised Lie derivative by non-derivative terms
\begin{equation}
\label{intro:L_def}
\widetilde{\mathbb{L}}_{\Lambda} = \mathbb{L}_{\Lambda} + \Lambda^{M} X_{M} \ ,
\end{equation}
where $\,X_M\,$ turns out to be $\,\En\,$ Lie algebra valued. In particular, it takes the form $\,(X_M)_N{}^P \equiv X_{MN}{}^P\,$ when acting on a field in the $\,\mathcal{R}_{\rm v}\,$ representation. Closure of the deformed generalised Lie derivative (\ref{intro:L_def}) and consistency of the tensor hierarchy require $\,X\,$ to be restricted to a specific \En representation (see Table~\ref{Table:intro_irreps}) and to satisfy a quadratic constraint
\begin{formula}
\label{intro:quadratic constraint}
	`X_MP^R `X_NR^Q - `X_NP^R `X_MR^Q + `X_MN^R `X_RP^Q  = 0  \ ,
\end{formula}
in analogy with the constraints appearing in gauged maximal supergravity \cite{Nicolai:2000sc,deWit:2007mt,deWit:2004nw,Bergshoeff:2007ef,Samtleben:2005bp,Puigdom:2008,FernandezMelgarejo:2011wx}.
Furthermore, an additional constraint involving both $\,X\,$ and $\,\partial_{M}\,$ must be imposed
\begin{eqnarray}
\label{intro:X-cond}
X_{MN}{}^{P} \,\, \partial_{P} &=& 0  \  .
\end{eqnarray}
This \,`$X$-constraint'\, can be interpreted as a compatibility condition between the $X$ deformation and the $\,y^{M}\,$ dependence of the fields and parameters. Together with the section constraint \eqref{intro:Y-cond} these conditions guarantee the consistency of the algebra of internal generalised diffeomorphisms and, ultimately, of the whole XFT.

\begin{table}[t!] 
\renewcommand{\arraystretch}{1.25}
\begin{center}
\scalebox{0.90}{
\begin{tabular}{|c|c|c|c|c|c|c|}
\hline 
$D$ & $9$ & $8$ & $7$ & $6$ & $5$ & $4$ \\[0.8mm]
\hline 
$\En$ & $\textrm{SL}(2) \times \mathbb{R}^{+}$ & $\textrm{SL}(2) \times \textrm{SL}(3)$ & $\textrm{SL}(5)$ & $\textrm{SO}(5,5)$ & $\textrm{E}_{(6)}$ & $\textrm{E}_{7(7)}$ \\[0.8mm]
\hline 
$\mathcal{R}_{\rm v}$ & $\textbf{2}_{3}+\textbf{1}_{-4}$ & $(\textbf{2},\textbf{3}')$ & $\textbf{10}'$ & $\textbf{16}_{c}$ & $\textbf{27}'$ & $\textbf{56}$ \\[0.8mm]
\hline 
$\mathcal{R}_{X}$ & $\textbf{2}_{-3}+\textbf{3}_{4}$ & $(\textbf{2},\textbf{3})+(\textbf{2},\textbf{6}')$ & $\textbf{15} + \textbf{40}'$ & $\textbf{144}_{c}$ & $\textbf{351}'$ & $\textbf{912}$ \\[0.8mm]
\hline
\end{tabular}}
\caption{Relevant $\,\En\,$ representations for the vector fields $\,A_{\mu}{}^{M}\,$ and the $X$ deformation \cite{Samtleben:2008pe}.}
\label{Table:intro_irreps} 
\end{center}
\end{table}

For specific choices of $\,X\,$, \eqref{intro:X-cond} is still compatible with solutions of the section constraint (\ref{intro:Y-cond}) that preserve \mbox{$\,n\,$} or $\,(n-1)\,$ internal coordinates.
The resulting XFT's ultimately describe three types of eleven- and ten-dimensional maximal supergravities:
\begin{itemize}
\item[$\circ$] $11$-dimensional and massless type IIA supergravities with background fluxes.
\item[$\circ$] Type IIB supergravity with background fluxes.
\item[$\circ$] Massive type IIA supergravity with background fluxes.
\end{itemize}
The latter case is a genuine result of XFT. Indeed, the massive IIA supergravity, which cannot be described in EFT without violating the section constraint, now admits a geometric description using the XFT framework. 
The background fluxes can be reabsorbed in the dynamical fields of the theory without violating the section constraint.
This is however not possible for the Romans mass.
As a result, the EGG of type IIA supergravity admits two inequivalent generalised Lie derivatives corresponding to the massless and massive theories, respectively\footnote{When it is non-vanishing, $m_{\rm R}$ can be rescaled by field redefinitions.}. It will also be shown how the XFT describing massive IIA can be related to a \textit{non-geometric} extension of EFT,  thus making contact with the DFT construction in \cite{Hohm:2011cp}.

The outline of the paper is as follows. In Section~\ref{sec:EFT} we review the main features of EFT and its generalised Lie derivative. From this formalism we reproduce the gauge transformations of the massless IIA theory and argue that a deformation of the generalised Lie derivative is needed in order to account for the gauge transformations in the massive case. In Section~\ref{sec:XFT} we present the general structure of the $X$ deformation, show that it contains the Romans mass parameter and classify deformations of the $\,\textrm{SL}(5)\,$ EFT compatible with ten- or eleven-dimensional solutions of the section- and $X$-constraints. In Section~\ref{sec:E7-XFT} we present the bosonic action, tensor hierarchy and transformation rules for the $\,\textrm{E}_{7(7)}\,$ XFT.
In Section~\ref{sec:non-geometry} we discuss the relation between XFT and a certain, possibly non-geometric, extension of EFT. We also comment on the construction of internal covariant derivatives.
We finally discuss some applications of our results in Section~\ref{sec:Discussion}. 


\section{Exceptional field theory and type IIA supergravity}
\label{sec:EFT}

Exceptional field theories (EFT's) embed the eleven-dimensional and massless type II supergravities in a unified framework, which renders the structure of their hidden exceptional symmetries manifest and captures the generalised geometries underlying them.
More concretely, the spacetime of eleven-dimensional supergravity is decomposed into a $D$ dimensional `external' spacetime and an $n=11-D$ dimensional `internal' space, without performing any truncation of degrees of freedom.
The internal diffeomorphisms are then extended to generalised diffeomorphisms accounting also for internal gauge transformations of the three- and six-form potentials (and of the dual graviton in $D=4,3$).
A similar situation occurs for the $D+(n-1)$ dimensional split of the massless type II supergravities. 
The set of internal coordinates is then extended to $\,y^M\,$, $\,M=1,\ldots,\dim{\cR_{\rm v}}\,$, to complete the representation of \En, which can be regarded as conjugate to the internal momenta and half-BPS charges of the theory \cite{AbouZeid:1999fv,Bossard:2015foa}.
A section constraint is imposed for consistency, restricting the coordinate dependence of fields and gauge parameters to a subset of the internal coordinates.
As long as one does not commit to a specific solution of this constraint, EFT can be regarded as being (formally) invariant under global $\En\times\bbR^+$ transformations.
The embedding of the original ten- and eleven-dimensional supergravities is recovered by choosing the appropriate solution of the section constraint. 
The generalised Lie derivative and other structures in EFT then reproduce (locally in a coordinate patch) the exceptional generalised geometry associated with the corresponding supergravity theory.

In their latest formulations, EFT's  have been constructed in any $D\ge3$ \cite{Hohm:2014fxa,Hohm:2013uia,Hohm:2013vpa,Abzalov:2015ega,Musaev:2015ces,Hohm:2015xna,Berman:2015rcc} following a prescription that mimics the structure of the maximal supergravities in the corresponding dimension \cite{Nicolai:2000sc,deWit:2007mt,deWit:2004nw,Bergshoeff:2007ef,Samtleben:2005bp,Puigdom:2008,FernandezMelgarejo:2011wx}.
In both the EFT's and the maximal supergravities, internal and spacetime symmetries completely specify the field content as well as its interactions in an elegant and unambiguous manner. In this section we introduce the basics of the EFT's which we will be extensively using.
We will focus on $D\ge4$ throughout this paper.

\subsection{Generalised diffeomorphisms}

EFT fields depend on spacetime coordinates $x^\mu$, $\mu=0,\ldots,D-1$, and extended internal coordinates $y^M$.
The fields and gauge parameters of the theory are arranged in objects that transform consistently under a set of exceptional generalised diffeomorphisms. On covariant objects generalised diffeomorphisms act with a generalised Lie derivative $\,\mathbb{L}_{\Lambda}\,$. 
For instance, the action of $\,\mathbb{L}_{\Lambda}\,$ on a vector $\,U^M\,$ of weight $\,\lambda(U)=\lambda_{U}\,$ reads\footnote{The transformation rule for a covariant tensor $V_M$ is deduced by requiring that the contraction $\,S=U^{M}V_{M}\,$ transforms as a scalar density of weight $\,\lambda_{U}+\lambda_{V}\,$.
The transformation rule for tensors follows immediately.} 
\begin{equation}
\label{def_Lie_1}
\mathbb{L}_{\Lambda} U^{M} = \Lambda^{N} \partial_{N} U^{M} - U^{N} \partial_{N} \Lambda^{M} + Y^{MN}{}_{PQ} \,\,  \partial_{N} \Lambda^{P} \,\, U^{Q} +  \, (\lambda_{U}-\omega) \, \partial_{P}\Lambda^{P} U^{M} ,
\end{equation}
where $\,\Lambda^M(x,y)\,$ is the gauge parameter, $\,Y^{MN}{}_{PQ}\,$ is a specific, constant $\En\times\bbR^+$ invariant tensor (so that $\,\delta_{\Lambda}Y^{MN}{}_{PQ}=\mathbb{L}_{\Lambda} Y^{MN}{}_{PQ}=0\,$), and $\,\omega=1/(D-2)\,$. All parameters of generalised diffeomorphisms carry weight $\,\omega\,$.

Consistency of the generalised diffeomorphisms requires the algebra of the generalised Lie derivative to close, namely
\begin{equation}
\label{eq:closure_EFT}
[ \mathbb{L}_{\Lambda} , \mathbb{L}_{\Sigma} ] W^{M}  =  \mathbb{L}_{[\Lambda,\Sigma]_{\textrm{E}}}W^{M} \ ,
\end{equation}
where the so-called E-bracket for parameters $\,\Lambda\,$ and $\,\Sigma\,$ is defined as
\begin{equation}
\label{eq:Ebracket}
\big[\Lambda,\Sigma\big]_{\textrm{E}}^M\,\equiv\,\frac12 \, (\mathbb{L}_\Lambda \Sigma^M-\mathbb{L}_{\Sigma} \Lambda^M)
\,=\, \Lambda^P \partial_P \Sigma^M + \frac12 \, Y^{MN}{}_{PQ} \, \partial_N\Lambda^P \, \Sigma^Q - (\Lambda \leftrightarrow \Sigma) \ .
\end{equation}
The requirement \eqref{eq:closure_EFT} translates into a set of conditions \cite{Berman:2012vc}
which severely restricts the dependence of the fields and parameters in the EFT on the generalised coordinates:
\begin{equation}
\label{sec_cond}
\hspace{-1.1mm}
\begin{array}{llll}
Y^{PQ}{}_{MN} \,\, \partial_{P} \otimes \partial_{Q} &=& 0 & , \\[2mm]
\big( Y^{M(P}{}_{TQ} \, Y^{T|N)}{}_{RS} - Y^{M(P}{}_{RS} \, \delta^{N)}_{Q} \big) \,\, (\partial_{P} \partial_{N})& = & 0 & , \\[2mm]
\big(Y^{MN}{}_{TQ} \, Y^{TP}{}_{[SR]} + 2 \, Y^{MN}{}_{[R|T|} \, Y^{TP}{}_{S]Q} - Y^{MN}{}_{[RS]} \, \delta^{P}_{Q} - 2 \, Y^{MN}{}_{[S|Q|} \, \delta^{P}_{R]} \big) \,\,  \partial_{(N} \otimes \partial_{P)}& = & 0 & , \\[2mm]
\big(Y^{MN}{}_{TQ} \, Y^{TP}{}_{(SR)} + 2 \, Y^{MN}{}_{(R|T|} \, Y^{TP}{}_{S)Q} - Y^{MN}{}_{(RS)} \, \delta^{P}_{Q} - 2 \, Y^{MN}{}_{(S|Q|} \, \delta^{P}_{R)} \big) \,\,  \partial_{[N} \otimes \partial_{P]}& = & 0 & .
\end{array}
\end{equation}
The first condition in (\ref{sec_cond}) is usually referred to as the \textit{section constraint}.
We will always impose that it holds on any combination of fields and/or parameters, including derivatives and products.
As a result, the section constraint restricts all objects in the EFT to depend only on a subset of the internal coordinates.
The other equations in \eqref{sec_cond} then follow from the section constraint for all the $\,\En\,$ EFT's \cite{Berman:2012vc}.

The E-bracket in (\ref{eq:Ebracket}) fails to satisfy the Jacobi identity:
\begin{equation}
\label{Jacobiator}
\big[[\Lambda,\Sigma]_\textrm{E},\Gamma\big]_\textrm{E}+\text{cycl.} = \frac13\,  \big\{[\Lambda,\Sigma]_\textrm{E},\Gamma\big\}_\textrm{E}+\text{cycl.} \ .
\end{equation}
This fact plays a central role in the construction of EFT's, as it requires the introduction of a hierarchy of $p$-form fields and gauge transformations \cite{Hohm:2013pua,Hohm:2013vpa,Hohm:2013uia,Hohm:2014fxa,Hohm:2015xna} similar to the one of gauged supergravities \cite{deWit:2005hv,deWit:2008ta}, in order to guarantee invariance of the equations of motion under generalised diffeomorphisms.
For vectors of weight $\omega$, one finds that the symmetric bracket $\left\{\Lambda,\Sigma\right\}_{\textrm{E}}\,$, reads 
\begin{equation}
\label{eq:symbracket_EFT}
\big\{\Lambda,\Sigma\big\}_\textrm{E}^M \,=\, \frac12 \,  (\mathbb{L}_\Lambda \Sigma^M+\mathbb{L}_\Sigma \Lambda^M) \,=\, \frac12 \, Y^{MN}{}_{PQ} \, \big[ \Sigma^Q\partial_N \Lambda^P+\Lambda^Q \partial_N \Sigma^P\big]  \ ,
\end{equation}
so that $\,\mathbb{L}_{\Lambda} \Sigma^M=\left[\Lambda,\Sigma\right]_\textrm{E}^M+\big\{\Lambda,\Sigma\}_\textrm{E}^M\,$.
Consistency of the EFT tensor hierarchy then follows from the fact that, upon using the section constraint, $\,\left\{\Lambda,\Sigma\right\}_{\textrm{E}}\,$ is a trivial gauge parameter, namely, $\,\mathbb{L}_{\{\Lambda,\Sigma\}_\textrm{E}}\,$ vanishes identically.

Covariance under internal generalised diffeomorphisms with parameters dependent on spacetime coordinates $\,x^\mu\,$ requires the introduction of appropriate covariant derivatives and associated connections \cite{Hohm:2013pua}
\begin{equation}
\partial_{\mu} \rightarrow \,\mathcal{D}_{\mu} \equiv \partial_{\mu}  - \mathbb{L}_{A_{\mu}} \ ,
\end{equation}
where $\,A_{\mu}{}^{M}(x,y)\,$ are the vector fields of EFT.
The requirement that $\,\cD_\mu\,$ is covariant fixes the transformation properties of $\,A_{\mu}{}^{M}\,$ up to the addition of trivial gauge parameters.
It is customary to choose
\begin{equation}
\label{deltaAM_1}
\delta_{\Lambda} A_{\mu}{}^{M} = \mathcal{D}_{\mu} \Lambda^{M} = \partial_{\mu} \Lambda^{M} - \mathbb{L}_{A_{\mu}} \Lambda^{M} \ .
\end{equation}
Making use of the fact that $\,\{\Lambda,A_\mu{}^M\}_{\rm E}\,$ is a trivial parameter, we can also give a different expression for $\,\delta_{\Lambda} A_{\mu}{}^{M}\,$ which will be convenient in the following section:
\begin{equation}
\label{deltaAM_2}
\delta_{\Lambda} A_{\mu}{}^{M} =  \partial_{\mu} \Lambda^{M} + \mathbb{L}_{\Lambda} A_{\mu}{}^{M} \ .
\end{equation}
The difference between any two choices of $\,\delta_{\Lambda} A_{\mu}{}^{M}\,$ is absorbed into the gauge transformations associated with the two-forms of the EFT tensor hierarchy.
The specifics of these tensor hierarchies depend on the dimension $\,D\,$, although a systematic treatment has been recently developed in \cite{Wang:2015hca}. 
We will discuss the $\,D=4\,$ case thoroughly in Section~\ref{sec:E7-XFT}.

\subsection{Massless IIA gauge transformations from EFT}
\label{MasslessIIA_EFT}

In order to make contact with the eleven-dimensional and massless type II supergravities, it is necessary to pick a specific solution of the section constraint in (\ref{sec_cond}). As preparation for the implementation of the Romans mass as a deformation parameter, here we will briefly exemplify how to recover the gauge transformations of ten-dimensional \textit{massless} IIA supergravity from those of EFT.

Let us start by introducing the massless gauge transformations of the IIA ten-dimensional \mbox{$p$-form} potentials $\,A_{\hat{M}}\,$, $\,A_{\hat{M}\hat{N}}\,$ and $\,A_{\hat{M}\hat{N}\hat{P}}\,$. These are specified by gauge parameters $\,\lambda\,$, $\,\Xi_{\hat{M}}\,$ and $\,\theta_{\hat{N}\hat{P}}=-\theta_{\hat{P}\hat{N}}\,$, where $\hat M,\,\hat N,\,\ldots$ are ten-dimensional spacetime indices, and take the form (we follow the conventions of ref.~\cite{Bandos:2003et})
\begin{equation}
\label{gauge_transfo_massless}
\delta A_{\hat{M}} = \partial_{\hat{M}} \lambda 
\hspace{5mm} , \hspace{5mm}
\delta A_{\hat{M}\hat{N}} = 2 \, \partial_{[\hat{M}} \,  \Xi_{\hat{N}]}
\hspace{5mm} , \hspace{5mm}
\delta A_{\hat{M}\hat{N}\hat{P}} = 3 \, \partial_{[\hat{M}} \,  \theta_{\hat{N}\hat{P}]} -  3 \, A_{[\hat{M}\hat{N}} \,  \partial_{\hat{P}]}  \lambda \ .
\end{equation}
For the sake of concreteness, we will consider a $7+3$ dimensional split of the fields and parameters of the ten-dimensional type IIA supergravity.
The $p$-forms of type IIA supergravity are decomposed in scalars, vectors and so on after appropriate Kaluza-Klein (KK) like redefinitions which are needed to achieve covariance under the seven-dimensional external diffeomorphisms.
All fields and gauge parameters still depend on the ten-dimensional coordinates $\,x^\mu\,$, $\,y^{\alpha}\,$ with $\,\mu=0, \ldots, 6\,$ and $\,\alpha=1,2,3\,$.
For instance the $D=7$ vectors arising from the ten-dimensional $p$-form potentials can be \mbox{written as}
\begin{equation}
\label{redef_KK}
A^{\textrm{KK}}_{\mu}  =   A_{\mu} - B_{\mu}{}^{\delta} \, A_{\delta}
\hspace{5mm} , \hspace{5mm}
A^{\textrm{KK}}_{\mu\beta}  =   A_{\mu \beta} - B_{\mu}{}^{\delta} \, A_{\delta \beta}
\hspace{5mm} , \hspace{5mm}
A^{\textrm{KK}}_{\mu \beta \gamma}  =   A_{\mu \beta \gamma} - B_{\mu}{}^{\delta} \, A_{\delta \beta \gamma} \ ,
\end{equation}
where $\,B_\mu{}^\alpha\,$ are the KK vector fields coming from the metric.
It is convenient to perform a second set of non-linear redefinitions\footnote{Similar redefinitions were discussed in refs~\cite{Godazgar:2013dma,Ciceri:2014wya,Guarino:2015vca}.}
\begin{equation}
\label{redef_non-linear}
C_{\mu}  =  A^{\textrm{KK}}_{\mu}
\hspace{8mm} , \hspace{8mm}
C_{\mu \beta}  =  A^{\textrm{KK}}_{\mu\beta} 
\hspace{8mm} \textrm{ and } \hspace{8mm}
C_{\mu \beta \gamma }  =  A^{\textrm{KK}}_{\mu \beta \gamma} +  A^{\textrm{KK}}_{\mu} \, A_{\beta \gamma}  \ .
\end{equation}
After some algebra manipulations it can be shown that, under \eqref{gauge_transfo_massless}, these vectors transform as follows under internal diffeomorphisms with parameter $\,\xi^\alpha\,$ and internal gauge transformations with parameters $\,\lambda\,$, $\,\Xi_\alpha\,$, $\,\theta_{\alpha\beta}\,$:
\begin{equation}
\label{Cvec_trans}
\begin{array}{llll}
\delta B_{\mu}{}^{\alpha} &=& (\partial_{\mu} - B_{\mu}{}^{\delta} \,\partial_{\delta})  \, \xi^{\alpha} + \xi^{\delta} \,\partial_{\delta}  B_{\mu}{}^{\alpha} \ , \\[1.5ex]
\delta C_{\mu} &=& \xi^{\delta} \,\partial_{\delta} C_{\mu}+ (\partial_{\mu} - B_{\mu}{}^{\delta} \,\partial_{\delta})  \, \lambda \ , \\[1.5ex]
\delta C_{\mu \beta} &=& \xi^{\delta} \,\partial_{\delta} C_{\mu \beta} + C_{\mu \delta} \, \partial_{\beta}\xi^{\delta}+ (\partial_{\mu} - B_{\mu}{}^{\delta} \, \partial_{\delta}) \,  \Xi_{\beta} + B_{\mu}{}^{\delta} \, \partial_{\beta} \,  \Xi_{\delta} \ , \\[1.5ex]
\delta C_{\mu \beta \gamma} &=& \xi^{\delta} \,\partial_{\delta} C_{\mu \beta \gamma} + 2 \, C_{\mu \delta [\gamma} \, \partial_{\beta]} \xi^{\delta} + (\partial_{\mu} - B_{\mu}{}^{\delta} \, \partial_{\delta})  \,  \theta_{\beta \gamma} + 2 \,   B_{\mu}{}^{\delta} \, \partial_{[\beta |} \,  \theta_{\delta  | \gamma]} & \\[.5ex]
&&+  2 \, C_{\mu} \,  \partial_{[\beta}  \Xi_{\gamma]}  -  2 \, C_{\mu [\beta} \,  \partial_{\gamma]}  \lambda \ .
\end{array}
\end{equation}

The $7+3$ dimensional split we have adopted to describe the massless IIA supergravity can be compared with the $D=7$ EFT, based on $\,{\rm E}_{4(4)}\equiv\SL(5)\,$ \cite{Blair:2013gqa,Berman:2010is,Musaev:2015ces}.
Analogous comparisons can be performed for other $\,D+(n-1)\,$ dimensional splits.
The \SL(5) EFT is characterised by generalised vectors $\Lambda^{M}$ in the $\bf{10}'$ representation, \textit{i.e.} $\,\Lambda^{mn}=-\Lambda^{nm}\,$, with $\,m=1,...,5\,$ being a fundamental SL(5) index. The structure tensor of the SL(5) EFT is given by\footnote{The entries in $\,Y^{mn \, pq}{}_{rs \, tu}$ are $0,\pm1\,$. Therefore, whenever an index pair $mn$ is contracted, a factor of $\tfrac{1}{2}$ must be explicitly included.}
\begin{equation}
\label{Y-tensor-SL5}
Y^{mn \, pq}{}_{rs \, tu} = \epsilon^{mnpqz} \, \epsilon_{rstuz} \ ,
\end{equation}
and the section constraint reduces to
\begin{equation}
\label{Y_cond_7D}
\epsilon^{mnpqz} \, \partial_{mn} \otimes \partial_{pq} = 0 \ .
\end{equation}
There are two inequivalent solutions of (\ref{Y_cond_7D}) (up to SL(5) transformations \cite{Blair:2013gqa}) corresponding to M-theory (more precisely, eleven-dimensional supergravity) and type IIB:
\begin{formula}
\label{sec_cond_7D}
\text{M-theory:}\qquad & \partial_{\alpha4}\neq0\,,\ \partial_{45}\neq0 \ \ \text{and}\ \ 
\partial_{\alpha5}=\partial_{\alpha\beta}=0\ ,\\[1ex]
\text{type IIB:}\qquad &\partial_{\alpha\beta}\neq0 \ \ \text{and}\ \
\partial_{\alpha4}=\partial_{\alpha5}=\partial_{45}=0\ .
\end{formula}
The massless IIA case is obtained by further restricting to only three coordinates in the M-theory solution. We will set $\,\partial_{45}=0\,$.

The SL(5) EFT contains $\textbf{10}'$ vector fields $A_{\mu}{}^M \equiv A_{\mu}{}^{mn}$ that transform under a generalised diffeomorphism as in (\ref{deltaAM_2}).
Using the massless IIA solution of the section constraint ($\partial_{\alpha4} \neq 0$), we can identify the field content and gauge parameters of the supergravity theory with those of the EFT:
\begin{equation}
\label{7D_identifications}
\begin{array}{lllllll}
A_{\mu}{}^{mn} &=& ( \, A_{\mu}{}^{\alpha 5} \, , \, A_{\mu}{}^{\alpha 4} \, , \,  A_{\mu}{}^{\alpha \beta} \, , \,  A_{\mu}{}^{45} \, )  
& = & ( \, \frac{1}{2} \, \epsilon^{\alpha \beta \gamma} \, C_{\mu \beta \gamma} \, , \, B_{\mu}{}^{\alpha} \, , \,   \epsilon^{\alpha \beta \gamma}  \, C_{\mu \gamma} \, , \,  C_{\mu} \, ) & , \\[2mm]
\Lambda^{mn} &=& ( \,  \Lambda^{\alpha 5} \, , \, \Lambda^{\alpha 4} \, , \,  \Lambda^{\alpha \beta} \, , \,  \Lambda^{45} \, )  
& = & ( \, \frac{1}{2} \, \epsilon^{\alpha \beta \gamma} \, \theta_{\beta \gamma} \,\,\,\,\, , \, \xi^{\alpha} \,\,\,\,\,\, , \,  \epsilon^{\alpha \beta \gamma}  \, \Xi_{\gamma} \,\,\,\, , \,  \lambda \, ) & , \\[2mm]
\partial_{mn} &=& ( \, \partial_{\alpha 5} \, , \, \partial_{\alpha 4} \, , \,  \partial_{\alpha \beta} \, , \,  \partial_{45} \, )  
& = & ( \, 0\,,\, \partial_{\alpha} \, , \, 0 \, , \,  0 \, , \,  0 \, ) \ .\end{array}
\end{equation}
After imposing the massless IIA solution of the section constraint, an explicit computation of the vector field transformations directly from (\ref{deltaAM_2}) reproduces \eqref{Cvec_trans}.
A similar analysis can be repeated for the other types of fields like the scalars or the two-  and three-form potentials.
However, the vector gauge transformations are enough for our purposes in the next section.

\subsection{Massive IIA gauge transformations from a deformed Lie derivative}
\label{MassiveIIA_XFT}

Let us now look at the gauge transformations of the ten-dimensional \textit{massive} IIA supergravity also in the $7+3$ dimensional split.
After performing the field redefinitions (\ref{redef_KK}) and (\ref{redef_non-linear}), the internal gauge transformations are modified by the Romans mass $m_{\rm R}$, yielding 
\begin{equation}
\label{Cvec_trans_massive}
\begin{array}{llll}
\delta B_{\mu}{}^{\alpha} &=& (\partial_{\mu} - B_{\mu}{}^{\delta} \,\partial_{\delta})  \, \xi^{\alpha} + \xi^{\delta} \,\partial_{\delta}  B_{\mu}{}^{\alpha} \ , \\[1.5ex]
\delta C_{\mu} &=& \xi^{\delta} \,\partial_{\delta} C_{\mu}+ (\partial_{\mu} - B_{\mu}{}^{\delta} \,\partial_{\delta})  \, \lambda - m_{\rm R} \,  B_{\mu}{}^{\delta} \, \Xi_{\delta} \ , \\[1.5ex]
\delta C_{\mu \beta} &=& \xi^{\delta} \,\partial_{\delta} C_{\mu \beta} + C_{\mu \delta} \, \partial_{\beta}\xi^{\delta}+ (\partial_{\mu} - B_{\mu}{}^{\delta} \, \partial_{\delta}) \,  \Xi_{\beta} + B_{\mu}{}^{\delta} \, \partial_{\beta} \,  \Xi_{\delta} \ , \\[1.5ex]
\delta C_{\mu \beta \gamma} &=& \xi^{\delta} \,\partial_{\delta} C_{\mu \beta \gamma} + 2 \, C_{\mu \delta [\gamma} \, \partial_{\beta]} \xi^{\delta} + (\partial_{\mu} - B_{\mu}{}^{\delta} \, \partial_{\delta})  \,  \theta_{\beta \gamma} + 2 \,   B_{\mu}{}^{\delta} \, \partial_{[\beta |} \,  \theta_{\delta  | \gamma]} \ \\[.5ex]
&&+  2 \, C_{\mu} \,  \partial_{[\beta}  \Xi_{\gamma]}  -  2 \, C_{\mu [\beta} \,  \partial_{\gamma]}  \lambda  - 2 \, m_{\rm R} \, C_{\mu [\beta} \, \Xi_{\gamma]} \ .
\end{array}
\end{equation}

Note that the extra terms in \eqref{Cvec_trans_massive} compared to \eqref{Cvec_trans} do not contain internal derivatives.
This poses an obstruction to recovering such variations from a standard EFT/generalised geometry Lie derivative like \eqref{def_Lie_1}, whose terms always contain derivatives of either the gauge parameter or the field it acts on.
However, the fact that massive IIA supergravity is a geometrically well-defined theory means that an exceptional generalised geometry describing it should still exist. This suggests that the solution to the above obstruction is to implement $\,m_{\rm R}\,$ as a deformation of $\,\bbL_\Lambda\,$, thus modifying the notion of covariance in the exceptional generalised geometry associated with type IIA supergravity.
The procedure we follow to deduce this deformation is the converse of what we discussed in the previous section: we still use the dictionary \eqref{7D_identifications} for the \SL(5) EFT, but we now repackage \eqref{Cvec_trans_massive} into an expression
\begin{equation}
\label{vec_transf_split_massive}
\delta_{\Lambda} A_{\mu}{}^{mn} = \partial_{\mu} \Lambda^{mn} + \widetilde{\mathbb{L}}_{\Lambda} A_{\mu}{}^{mn} \ ,
\end{equation}
where $\,\wtd\bbL_\Lambda\,$ accounts for $\,m_{\rm R}\,$ and reduces to the standard EFT Lie derivative in the limit ${\,m_{\rm R}\to0}\,$.
We stress that vector fields transform faithfully under internal generalised diffeomorphisms, so that by covariance this procedure uniquely identifies the deformation induced by $\,m_{\rm R}\,$ for every other field, too.
The resulting deformed Lie derivative reads
\begin{equation}
\label{DeformL}
\widetilde{\mathbb{L}}_{\Lambda} A_{\mu}{}^{mn} = \mathbb{L}_{\Lambda} A_{\mu}{}^{mn} - X_{pq \,  rs}{}^{mn} \, \Lambda^{pq} \, A_{\mu}{}^{rs} \ ,
\end{equation}
where the second term in the r.h.s of (\ref{DeformL}) is specified by an $X$ deformation of the form
\begin{align}
\label{X6indices}
X_{mn \, pq}{}^{rs} &= 2 \, X_{mn \, [p}{^{[r}} \, \delta_{q]}^{s]} \ ,
\intertext{with non-vanishing entries given by}
\label{X4indices}
X_{\alpha \beta \, \gamma}{}^{5} &= -2 \, m_{\rm R} \, \epsilon_{\alpha \beta \gamma} \ ,
\end{align}
and where $\,\epsilon_{\alpha \beta \gamma}\,$ is the Levi-Civita symbol in three dimensions with $\epsilon_{123}=+1\,$.  
Note at this point that equations (\ref{X6indices}) and (\ref{X4indices}) correspond to the embedding tensor of the gauged maximal supergravity induced by a three-torus compactification of massive IIA supergravity\footnote{The reduced theory is a seven-dimensional gauged maximal supergravity with three vectors $\,A_{\mu}{}^{\alpha \beta}\,$ spanning an abelian $\,\mathbb{R}^{3}\,$ gauging specified by the three commuting generators $\,t^{\gamma}{}_{5}\,$.}.

Consistency requirements like closure of $\,\wtd\bbL\,$ will follow from consistency of the original massive IIA theory, \emph{at least} as long as we restrict to the solution of the section constraint that corresponds to the type IIA theory.
As we shall see, however, the structures unveiled in this section can be immediately generalised to other dimensions as well as to generic $X$ deformations.
Therefore we will discuss consistency of the deformed EFT's in a more general setting in the next section, to later come back to the case of the Romans mass.

\section{Deformations of exceptional field theory}
\label{sec:XFT}

Motivated by the Romans mass deformation of the SL(5) EFT found in the previous section, we move to investigate general deformations of EFT. In this section we will focus on the structure of generalised diffeomorphisms and discuss their closure and consistency conditions.

\subsection{Some notions of gauged maximal supergravity}
\label{sec:Emb_tens}

It will be useful for our purposes to first review a few basic aspects of the embedding tensor formalism of gauged maximal supergravities.
An incomplete list of references dealing with gauged maximal supergravities in $\,d=4,7,9\,$ dimensions includes refs~\cite{Samtleben:2005bp,deWit:2007mt,FernandezMelgarejo:2011wx}. 

The gauge group of a gauged maximal supergravity in $\,D\,$ dimensions must be a subgroup of  $\,\En\,$, where $\,n=11-D\,$.
This is the global exceptional symmetry of the ungauged theory.
We will exclude in our discussion the gauging of the $\,\bbR^+\,$ trombone symmetry of maximal supergravities \cite{LeDiffon:2008sh,LeDiffon:2011wt}.  
The supergravity Lagrangian and symmetry variations are entirely specified by an \emph{embedding tensor} $\,`\Theta_M^{\alpha} \,$, where $\,\alpha\,$ is an $\,\En\,$ adjoint index and $\,M\,$ is in the $\,\cR_{\rm v}\,$ representation.
Equivalently, introducing $\,\En\,$ generators $\,[t_{\alpha}]_{M}{}^{P}\,$, we can construct an object with $\,\cR_{\rm v}\,$ indices only
\begin{equation}
\label{X to Theta}
	X_{MN}{}^{P} =  \Theta_{M}{}^{\alpha} \, [t_{\alpha}]_{N}{}^{P} \ ,
\end{equation}
which captures the same information as $\,`\Theta_M^{\alpha} \,$.\footnote{This is not necessarily true for non-maximal theories.}

Despite its name, the embedding tensor should be taken as a fixed object, which therefore explicitly breaks the global $\,\En\,$ symmetry in order to gauge a subgroup.
Its variations under diffeomorphisms, gauge transformations and supersymmetry all vanish by definition.
However, it is often convenient to regard $\,`X_MN^P \,$ as a spurious object, which is allowed to transform under $\,\En\,$ together with the fields of the theory, thus obtaining a formally $\,\En\,$ covariant treatment of gauged maximal supergravity.
A set of quadratic and linear constraints must be imposed on the embedding tensor for consistency of the gauged theory.
The quadratic one
\begin{formula}
\label{quadratic constraint}
	`X_MP^R `X_NR^Q - `X_NP^R `X_MR^Q + `X_MN^R `X_RP^Q  = 0  \ ,
\end{formula}
ensures closure of the gauge algebra and requires that the embedding tensor, when regarded as a spurious object, is invariant under the gauge transformations it defines. The linear constraint is required both by supersymmetry and, at the bosonic level, by imposing that the hierarchy of $p$-form fields induced by the gauging is consistent with the representation content and counting of degrees of freedom of the ungauged theory.
In practice, the linear constraint restricts the embedding tensor to specific irreps (denoted by $\,\cR_X\,$ in Table~\ref{Table:intro_irreps}) contained in the tensor product of $\,\cR_{\rm v}\,$ and the adjoint representation
\begin{formula}
\label{linear constraint general}
	\Theta\in \cR_X\ \subset\ \cR_{\rm v}\otimes\mathrm{adj}  \ .
\end{formula}

\subsection{Deformed generalised Lie derivative}
\label{sec:Gen_diff_X}

Motivated by our discussion of the internal gauge variations of massive IIA supergravity, we will now consider generic deformations of the exceptional generalised Lie derivative $\,\bbL_\Lambda\,$ of the $D$-dimensional EFT by non-derivative terms specified by a constant object $\,X_{MN}{}^{P}\,$.
As we will see, this object satisfies the same requirements as the embedding tensor of the $D$-dimensional gauged maximal supergravity:
the quadratic constraint \eqref{quadratic constraint} arises from the closure and Jacobi identity for the generalised diffeomorphisms, while the linear or representation constraint (see Table~\ref{Table:intro_irreps}) is required for consistency of the resulting hierarchy of tensor fields.
We exclude deformations of the trombone type from our discussion.

We thus start by introducing a deformed generalised Lie derivative which acts on vectors as
\begin{equation}
\label{def_Lie_2}
\widetilde{\mathbb{L}}_{\Lambda} U^{M} = \mathbb{L}_{\Lambda} U^{M} - X_{NP}{}^{M} \, \Lambda^{N} \, U^{P} \ , 
\end{equation}
where the standard (undeformed) generalised Lie derivative $\,\bbL_\Lambda\,$ is defined in (\ref{def_Lie_1}).
A first consistency requirement is that $\,\widetilde{\mathbb{L}}_{\Lambda}\,$ is compatible with the global $\,\En\,$ structure of the theory: thus $`X_MN^P $ must decompose just as in \eqref{X to Theta}.
We can thus say that in general
\begin{equation}
\label{eq:def}
\widetilde{\mathbb{L}}_{\Lambda} \equiv \mathbb{L}_{\Lambda} + \Lambda^{M} X_{M} \ ,
\end{equation}
where $X_M$ is $\,\En\,$ Lie algebra valued and acts in the appropriate representation.

Closure of generalised diffeomorphisms translates in the deformed version of \eqref{eq:closure_EFT}:
\begin{equation}
\label{eq:closure}
\big[\widetilde{\mathbb{L}}_\Lambda,\widetilde{\mathbb{L}}_\Sigma\big] = \widetilde{\mathbb{L}}_{[\Lambda,\Sigma]_X} \ ,
\end{equation}
where the $\,X$-bracket $\,[\cdot,\cdot]_X\,$ takes the form
\begin{equation}
\label{eq:Xbracket}
\big[\Lambda,\Sigma\big]_{X}^M \equiv \frac12 \, (\widetilde{\mathbb{L}}_\Lambda \Sigma^M - \widetilde{\mathbb{L}}_{\Sigma} \Lambda^M) = \big[\Lambda,\Sigma\big]_{\textrm{E}}^M   - X_{[PQ]}{}^{M}\, \Lambda^{P} \, \Sigma^{Q} \ .
\end{equation}
Requiring \eqref{eq:closure} induces a new set of consistency constraints.
Since $\,\Lambda\,$ and $\,\Sigma\,$ are arbitrary parameters, these constraints can be separated based on the number of derivatives. The two-derivative ones do not contain $\,X_M\,$ and therefore reduce to the original section constraint \eqref{sec_cond}. 
An explicit computation yields
\begin{formula}
\label{commutator}
[ \widetilde{\mathbb{L}}_{\Lambda} , \widetilde{\mathbb{L}}_{\Sigma} ] W^{M}  -  \widetilde{\mathbb{L}}_{[\Lambda,\Sigma]_{X}}W^{M} =\ & 
A^{M}_{NPS} \,\,  \Lambda^{N} \Sigma^{P} W^{S} 
+ X_{[NP]}{}^{Q} \,\, \Lambda^{N} \Sigma^{P} \partial_{Q}W^{M}  \\[1ex]
& + B^{MQ}_{NRS} \,\,  (\Lambda^{N} \partial_{Q}\Sigma^{R} W^{S} - \partial_{Q}\Lambda^{R} \Sigma^{N} W^{S}) \ ,
\end{formula}
where, without loss of generality, we have already assumed  \eqref{sec_cond} to hold.
The r.h.s. of (\ref{commutator}) therefore defines $X$-dependent conditions.
The $A$ and $B$ terms read
\begin{formula}
\label{AB_tensors}
A^{M}_{NPS} =\ &  2 X_{[N|Q}{}^{M} X_{P]S}{}^{Q} - X_{QS}{}^{M} X_{[NP]}{}^{Q} \ , \\[2ex]
B^{MQ}_{NRS} =\ &  X_{(NR)}{}^{M} \delta_{S}^{Q} - X_{NS}{}^{Q} \delta_{R}^{M} \\
&+ Y^{MQ}{}_{RP} X_{NS}{}^{P} - Y^{PQ}{}_{RS} X_{NP}{}^{M} + Y^{MQ}{}_{PS} X_{[NR]}{}^{P} - \frac{1}{2} Y^{PQ}{}_{RN} X_{PS}{}^{M} \ .
\end{formula}
Note that the first line is the antisymmetric part of the quadratic constraint \eqref{quadratic constraint}.
Altogether, we have the requirements
\begin{formula}
\label{closure eqs}
	 A^{M}_{NPS}=0 \hspace{5mm} , \hspace{5mm}
	`X_[NP]^Q \, \partial_Q = 0 \hspace{5mm} \text{ and } \hspace{5mm}
	 B_{NRS}^{MQ} \, \partial_Q = 0\ .
\end{formula}

The conditions above are not yet final.
Just as for the E-bracket, the $\,X$-bracket fails to define a Lie algebra as the Jacobi identity does not hold.
Instead, it yields a  Jacobiator
\begin{equation}
\big[[\Lambda,\Sigma]_X,\Gamma\big]_X+\text{cycl.} \,\, = \frac13 \, \big\{[\Lambda,\Sigma]_X,\Gamma\big\}_X+\text{cycl.}\, \ , 
\end{equation}
where the $X$-modified symmetric bracket turns out to be
\begin{equation}
\label{eq:symbracket}
\big\{\Lambda,\Sigma\big\}_X^M \equiv \frac12 (\widetilde{\mathbb{L}}_\Lambda \Sigma^M+\widetilde{\mathbb{L}}_\Sigma \Lambda^M)= \big\{\Lambda,\Sigma\big\}_\textrm{E}^M - X_{(PQ)}{}^M  \Lambda^P  \Sigma^Q \ .
\end{equation}
Consistency of the XFT requires that the Jacobiator again corresponds to a trivial gauge parameter, namely, $\,\widetilde{\mathbb{L}}_{\{\Lambda,\Sigma\}_{X}}\,$ vanishes.
A direct computation shows that
\begin{align}
\label{Trivial_X_equation}
\widetilde{\mathbb{L}}_{\{\Lambda,\Sigma\}_{X}} U^{M} =\ &   C^{MR}_{SPQ} \,\,  ( \Lambda^{Q} \partial_{R}\Sigma^{P} U^{S} + \partial_{R} \Lambda^{P} \Sigma^{Q} U^{S})  
 -  X_{(PQ)}{}^{R} \,\, \Lambda^{P} \Sigma^{Q} \, \partial_{R}U^{M} \\[1ex]\nonumber
& +  X_{(PQ)}{}^{R} \, X_{RS}{}^{M}\, \Lambda^{P} \Sigma^{Q} U^{S} \ ,%
\shortintertext{with}%
\label{C_tensor}
C^{MR}_{SPQ} =\ & X_{(PQ)}{}^{M} \delta_{S}^{R}  - Y^{MR}{}_{TS} \, X_{(PQ)}{}^{T} - \frac{1}{2} \,  Y^{TR}{}_{PQ} \, X_{TS}{}^{M} \ ,
\end{align}
and where we have used the conditions (\ref{closure eqs}) derived from \eqref{commutator}.
Therefore we must impose
\begin{equation}
\label{Trivial_X}
X_{(PQ)}{}^{R} \, X_{RS}{}^{M} = 0 \hspace{5mm} , \hspace{5mm}
X_{(PQ)}{}^{R} \,\, \partial_{R} = 0 \hspace{5mm} \text{ and } \hspace{5mm}
C^{MR}_{SPQ} \,\,  \partial_{R} = 0\ .
\end{equation}
The first equation in \eqref{Trivial_X} combines with the first equation in \eqref{closure eqs} to produce the full set of quadratic constraints in \eqref{quadratic constraint}.
The middle equations in \eqref{closure eqs} and \eqref{Trivial_X} combine into the \mbox{$X$-constraint} $\,\,`X_MN^P \, \partial_P = 0\,$.
A careful analysis of the representation content of the remaining conditions (namely, the `$B$' and `$C$' terms) shows that they are entirely equivalent to the \mbox{$X$-constraint}.
We thus arrive at the final set of consistency conditions for the deformed generalised Lie derivative \eqref{eq:def}:
\begin{align}
\label{section constraint}
`Y^MN_PQ \,\, \partial_M \otimes \partial_N &= 0
\qquad\text{( section constraint )}\ ,
\\[1ex]
\label{X condition}
`X_MN^P \, \partial_P &= 0
\qquad\text{( $X$-constraint )}
\ ,
\end{align}
and $`X_M $ must additionally satisfy the quadratic constraint in \eqref{quadratic constraint}.
The above conditions should be intended as acting on any field, parameter and combinations thereof.
As a result, the new $X$-constraint \eqref{X condition} restricts the coordinate dependence to those coordinates left invariant by the $\,\En\,$ elements generated by $\,X_M\,$.
Together with the linear and quadratic constraints on $X$, this is the only new condition required for consistency of the deformed EFT.

We should also emphasise that our notion of covariance under internal generalised diffeomorphisms is now given in terms of $\,\widetilde\bbL\,$, so that $\,\delta_\Lambda T = \widetilde\bbL_\Lambda T\,$ for any tensor $\,T\,$.
The deformation $\,`X_MN^P \,$ by definition does not vary under any (internal or external) diffeomorphism and gauge transformations.
Its generalised Lie derivative, instead, does not necessarily vanish.
Using the constraints above one can compute
\begin{equation}
\label{deltaX}
\widetilde{\mathbb{L}}_{\Lambda} X_{MN}{}^{P} =
 2 \, \partial_{[M} \, \Lambda^{R} \, X_{|R|N]}{}^{P}  + Y^{PQ}{}_{RN} \, \partial_{Q}\Lambda^{S} \, X_{SM}{}^{R}\ ,
\end{equation}
where we assign the weight $\lambda(X) = -\omega$, as can be deduced by requiring that generalised Lie derivatives of tensors maintain a definite weight.

We close the section by stressing again that $\,`X_MN^P \,$ is restricted to the $\,\En\,$ representations displayed in Table~\ref{Table:intro_irreps} for consistency of the tensor hierarchy.

\subsection{Section constraint and massive IIA supergravity}
\label{subsec:SC}

Equipped with the new generalised Lie derivative $\,\widetilde{\mathbb{L}}\,$ and consistency conditions derived in the previous section, we now look at specific $X$ deformations to discuss their interpretation.
We will come back to the construction of the full XFT action in Section~\ref{sec:E7-XFT} where we discuss the $\,\Eseven\,$ case in detail.
 Starting from the M-theory solution of the section constraint (\ref{section constraint}), we now show how turning on the $X$ deformation corresponding to the Romans parameter~$\,m_{\rm R}\,$, to which we refer as $\,\XR\,$, proves no longer compatible with a dependence of the fields and parameters on the M-theory coordinate as a consequence of the $X$-constraint (\ref{X condition}). The resulting XFT will then describe the massive IIA theory. 
The deformation $\XR$ always corresponds to the embedding tensor obtained from reduction of massive IIA supergravity on a torus.
We will also show that other solutions of the section and $X$-constraints compatible with $\XR$ exist and correspond to type II theories with background RR $p$-form fluxes T-dual to $\,m_{\rm R}\,$. For the $\,D=4\,$ case, we will also find an eleven-dimensional supergravity solution.
Several of these solutions would be equivalent to each other in standard EFT, as they belong to the same $\,\En\,$ orbit.
However, the presence of $\XR$ breaks $\,\En\,$ to a subgroup which always contains at least an $\,\SL(n-1)\,$ factor, and solutions to the constraints must be classified in orbits of this subgroup only.

\subsubsection*{$\textrm{SL}(2)\times \mathbb{R}^{+}$ XFT}

The EFT with $\,(D,n)=(9,2)\,$ features an $\,\textrm{SL}(2)\times \mathbb{R}^{+}\,$ structure and has recently been constructed in ref.~\cite{Berman:2015rcc}. The extended space has coordinates $\,y^{M}=(y^{\alpha} \,,\, y^{3})\,$ with $\,\alpha=1,2\,$ being a fundamental $\,\textrm{SL}(2)\,$ index. The $\,\textrm{SL}(2)\times \mathbb{R}^{+}\,$ invariant $Y$-tensor is given by
\begin{equation}
Y^{\alpha 3}{}_{\beta 3} = Y^{\alpha 3}{}_{3 \beta} = Y^{3 \alpha}{}_{\beta 3} = Y^{3 \alpha}{}_{3 \beta} = \delta^{\alpha}_{\beta} \ ,
\end{equation}
and the section constraint in (\ref{section constraint}) reduces to $\,\partial_{\alpha} \otimes \partial_{3} = 0\,$. There are two inequivalent solutions corresponding to M-theory and type IIB supergravity
\begin{equation}
\label{sec_cond_9D}
i) \,\,\,\,\, \partial_{\alpha} \neq 0 \,\,\,\, , \,\,\,\, \partial_{3} = 0  \,\,\,\,\, \textrm{(M-theory)} 
\hspace{8mm} \textrm{and } \hspace{8mm} 
ii) \,\,\,\,\,  \partial_{3} \neq 0 \,\,\,\, , \,\,\,\, \partial_{\alpha} = 0   \,\,\,\,\, \textrm{(type IIB)}  \ .
\end{equation}

In the context of maximal $\,D=9\,$ supergravity \cite{Bergshoeff:2002nv,Nishino:2002zi,FernandezMelgarejo:2011wx}, the Romans mass parameter induces an embedding tensor\footnote{The corresponding gauging is simply a shift symmetry $\,\mathbb{R}\,$ generated by $\,t^{2}{}_{1} \in \textrm{SL}(2)$ and spanned by the vector field $\,A_{\mu}{}^{3}\,$ \cite{Bergshoeff:1996ui,Bergshoeff:2002nv}.}  with only non-vanishing entry $\,[\XR]_{3 2}{}^{1} = m_{\rm R}\,$. Taking it to be the $X$~deformation in XFT and substituting it into the $X$-constraint (\ref{X condition}) yields
\begin{equation}
\label{new_sec_cond_9D}
\begin{array}{llll}
m_{\rm R} \, \partial_{1} = 0 \ .
\end{array}
\end{equation}
As a result, any dependence on the M-theory coordinate  $\,y^{1}\,$ is removed by the $X$-constraint (\ref{new_sec_cond_9D}) reflecting the fact that massive IIA cannot be embedded into M-theory. Using (\ref{new_sec_cond_9D}) to simplify the section constraint in (\ref{section constraint}) one finds $\,\partial_{2}\otimes \partial_{3} = 0\,$, which gives rise to the type IIA and IIB solutions
\begin{equation}
\label{sec_cond_9D_massive}
i) \,\,\,\,\, \partial_{2} \neq 0 \,\,\,\, , \,\,\,\, \partial_{3} = 0 \,\,\,\,\, \textrm{(type IIA)}
\hspace{8mm} \textrm{and } \hspace{8mm} 
ii) \,\,\,\,\, \partial_{3} \neq 0 \,\,\,\, , \,\,\,\, \partial_{2} = 0 \,\,\,\,\, \textrm{(type IIB)} \ .
\end{equation}

In the IIA solution, the $\XR$ deformation is identified with the Romans mass. In the IIB solution, the same $\XR$ deformation corresponds to turning on a RR background flux $\,F_{(1)}\,$ along the $\,y^{3}\,$ coordinate. The two solutions are related by a T-duality transformation 
\begin{equation}
 \,\,  i) \,\, \textrm{massive IIA} \,\,  \overset{T}{\longrightarrow}
 \,\,  ii) \,\, \textrm{IIB with } F_{(1)}  \ ,
\end{equation}
exchanging $\,y^{2} \leftrightarrow y^{3}\,$.

\subsubsection*{$\textrm{SL}(5)$ XFT}

The EFT with $\,(D,n)=(7,4)\,$ possesses an $\,\textrm{SL}(5)\,$ structure and has already been discussed in Section~\ref{MasslessIIA_EFT}. The $\,\textrm{SL}(5)\,$ invariant $Y$-tensor and the section constraint, as well as its solutions, can be found in (\ref{Y-tensor-SL5}), (\ref{Y_cond_7D}) and (\ref{sec_cond_7D}). The consistent $\XR$ deformation induced by the Romans mass was presented in Section~\ref{MassiveIIA_XFT}, resulting in eqs~(\ref{X6indices})  and (\ref{X4indices}). The  $X$-constraint (\ref{X condition}) reads in this case
\begin{equation}
\label{new_sec_cond_7D}
\begin{array}{llll}
 m_{\rm R} \, \partial_{\alpha 5} = m_{\rm R} \, \partial_{45}= 0  \ .
\end{array}
\end{equation}
Note again that any dependence on the M-theory coordinate $\,y^{45}\,$ as well as on the `brane coordinates' $\,y^{\alpha 5}\,$ is killed by the $X$-constraint (\ref{new_sec_cond_7D}).
Substituting (\ref{new_sec_cond_7D}) in the section constraint (\ref{section constraint}) produces $\,\epsilon^{\alpha \beta \gamma} \partial_{\alpha 4} \otimes \partial_{\beta \gamma} = 0\,$, which gives rise to the `natural' type IIA and IIB solutions
\begin{equation}
\label{sec_cond_7D_massive}
i) \,\, \partial_{\alpha 4} \neq 0 \,\,\,\, , \,\,\,\, \partial_{\alpha \beta} = 0  \,\,\,\, \textrm{(type IIA)}
\hspace{8mm} \textrm{ and } \hspace{8mm} 
iv) \,\, \partial_{\alpha \beta} \neq 0 \,\,\,\, , \,\,\,\, \partial_{\alpha 4}=0   \,\,\,\, \textrm{(type IIB)} \ ,
\end{equation}
together with two more solutions  (with $\alpha \neq \beta \neq \gamma$)
\begin{equation}
\label{sec_cond_7D_hybrid}
\begin{array}{lllll}
ii) & \partial_{\alpha 4} \, , \,  \partial_{\beta 4} \, , \,  \partial_{\alpha \beta} \neq 0   \,\,\,\, , \,\,\,\, \partial_{\gamma 4} =\partial_{\beta \gamma} =\partial_{\gamma \alpha} = 0   &\textrm{(type IIB)} & , \\[2mm]
iii) & \partial_{\alpha 4} \, , \,  \partial_{\alpha \beta} \, , \,  \partial_{\gamma \alpha} \neq 0   \,\,\,\, , \,\,\,\, \partial_{\beta 4} =\partial_{\gamma 4} =\partial_{\beta \gamma} = 0 & \textrm{(type IIA)} & .
\end{array}
\end{equation}

In the IIA solution $i)$, the $\XR$ deformation is identified with the Romans mass. In the IIB solution $ii)$, it corresponds to a RR background flux $\,F_{(1)}\,$ along the single coordinate~$\,y^{\alpha \beta}\,$. In the IIA solution $iii)$, it maps to a RR background flux $\,F_{(2)}\,$ along the two coordinates  $\,(y^{\alpha \beta},y^{\gamma \alpha})\,$. Finally, in the IIB solution $iv)$, the $X$ deformation corresponds to a RR background flux $\,F_{(3)}\,$. The four solutions are connected via a chain of T-duality transformations
\begin{equation}
 \,\,  i) \,\, \textrm{massive IIA} \,\,  \overset{T_{\gamma}}{\longrightarrow}
 \,\,  ii) \,\, \textrm{IIB with } F_{(1)} \,\,  \overset{T_{\beta}}{\longrightarrow}
 \,\,  iii) \,\, \textrm{IIA with } F_{(2)} \,\,  \overset{T_{\alpha}}{\longrightarrow}
 \,\,  iv) \,\, \textrm{IIB with } F_{(3)} \ ,
\end{equation}
where $T_{\gamma}$ exchanges $y^{\gamma 4} \leftrightarrow y^{\alpha \beta}$,  with $\alpha \neq \beta \neq \gamma$.

\subsubsection*{$\textrm{E}_{7(7)}$ XFT}

The EFT with $\,(D,n)=(4,7)\,$ features an $\,\textrm{E}_{7(7)}\,$ structure and the coordinates~$\,y^{M}\,$ of the extended space transform in the $\,\bf{56}\,$ fundamental representation. The $\,\textrm{E}_{7(7)}\,$ invariant $Y$-tensor reads \cite{Berman:2012vc}
\begin{equation}
\label{Y-tensor_4D}
Y^{MN}{}_{PQ}=-12 \, [t_\alpha]^{MN}[t^\alpha]_{PQ}-\frac12 \, \Omega^{MN} \, \Omega_{PQ} \ ,
\end{equation}
where $\,[t_\alpha]_{M}{}^{N}\,$ are the E$_{7(7)}$ generators. Fundamental indices are raised and lowered using the Sp(56)-invariant (and thus $\textrm{E}_{7(7)}$-invariant) antisymmetric tensor $\,\Omega_{MN}\,$.\footnote{We use the NW-SE conventions of ref.~\cite{Hohm:2013uia} such that $\,[t_\alpha]^{MN} =\Omega^{MP} [t_{\alpha}]_{P}{}^N\,$, $\,[t_{\alpha}]_{MN} = [t_{\alpha}]_{M}{}^P \, \Omega_{PN}\,$ and $\,\Omega^{MP}\Omega_{NP} = \delta_N^M\,$.} 
It will prove convenient to move to an SL(8)-covariant description of the theory where one has the $\textrm{E}_{7(7)} \supset \textrm{SL}(8)$ branching $\textbf{56} \rightarrow \textbf{28}' + \textbf{28}$. For instance, the coordinates $\,y^{M}=(y^{AB} \, , \, y_{AB})\,$ are expressed in terms of an antisymmetric pair $AB$ of SL(8) fundamental indices $\,A,B=1,...,8\,$.
The section constraint (\ref{section constraint}) reads
\begin{equation}
\label{Y_cond_4D_SL8}
\begin{array}{rlll}
\partial_{[AB}  \otimes \partial_{CD]} - \dfrac{1}{4!} \, \epsilon_{ABCD EFGH} \, \partial^{EF}  \otimes \partial^{GH}& = &  0 & , \\[2mm]
\partial_{AC} \otimes \partial^{BC} + \partial^{BC} \otimes \partial_{AC}  - \dfrac{1}{8} \,  \delta_{A}^{B} \, \left( \partial_{CD} \otimes \partial^{CD} + \partial^{CD} \otimes \partial_{CD} \right)  &=& 0 & , \\[2mm]
 \partial_{CD} \otimes \partial^{CD} - \partial^{CD} \otimes \partial_{CD} & = &  0 & .
\end{array}
\end{equation}
Branching now the SL(8) index with respect to $\,\textrm{SL}(7) \subset \textrm{SL}(8)\,$, namely $\,A=(I,8)\,$ with $\,{I=1,...,7}\,$, two solutions of (\ref{Y_cond_4D_SL8}) were identified in \cite{Hohm:2013uia} (see also the discussion in Section~3.2 of ref.~\cite{Baron:2014yua}) which are the ordinary \mbox{M-theory} and type IIB solutions\footnote{As a representative of the IIB solutions, we pick the one that is obtained by acting with three T-dualities upon the `natural' IIA solution (which follows from the M-theory one after imposing $\partial_{78}=0$).}. These are the only maximal solutions up to $\,\textrm{E}_{7(7)}\,$ transformations \cite{Bossard:2015foa,Bandos:2015rvs} and  involve a non-trivial dependence on the extended space of the form
\begin{equation}
\label{sec_cond_4D}
i) \,\, \partial_{I8} \neq 0  \,\,\,\,\,\, \textrm{(M-theory)} 
\hspace{10mm} \textrm{ and } \hspace{10mm}
ii) \,\, \partial_{\alpha 8} \neq 0 \,\,\,\,\, , \,\,\,\,\, \partial^{\hat{\alpha} 7} \neq 0   \,\,\,\,\,\, \textrm{(type IIB)} \  ,
\end{equation}
where we have further split $\,I=(m,7)\,$ and  $\,m=(\alpha , \hat{\alpha})\,$ with $\,m=1,...,6\,$, $\,\alpha=1,2,3\,$ and $\,\hat{\alpha}=4,5,6\,$. 

In the context of maximal $\,D=4\,$ supergravity \cite{deWit:2007mt}, the Romans mass induces a consistent embedding tensor of the form\footnote{The induced gauging in four dimensions is an abelian $\,\mathbb{R}^{7}\,$ symmetry associated to the generators $\,t^{I}{}_{8} \in \textrm{SL}(8)\,$ and it is spanned by the vector fields $\,A_{\mu \, I8}\,$ \cite{Guarino:2015qaa}.}
\begin{equation}
\label{X_4D_Romans}
[{\XR}]^{AB \phantom{CD}EF}_{\phantom{AB}CD} = - {[{\XR}]^{AB \,  EF}}_{CD} = -8 \, \delta_{[C}^{[A} \xi^{B][E} \delta_{D]}^{F]} \ ,
\end{equation}
with $\,\xi^{AB}= m_{\rm R} \, \delta^{A}_{8} \, \delta^{B}_{8}\,$. Taking  now (\ref{X_4D_Romans}) to be the $X$ deformation in XFT produces an $X$-constraint (\ref{X condition}) of the form 
\begin{equation}
\label{new_sec_cond_4D}
\begin{array}{llll}
 m_{\rm R} \, \partial_{I 8} = m_{\rm R} \, \partial^{IJ}= 0  \ ,
\end{array}
\end{equation}
which removes any dependence on the M-theory coordinate $\,y^{78}\,$ as well as on the `brane coordinates' $\,y^{m 8}\,$ and $\,y_{IJ}\,$. Substituting (\ref{new_sec_cond_4D}) in the section constraint (\ref{section constraint}) reduces it to two conditions $\,\partial_{[IJ}   \otimes    \partial_{KL]}  = 0 \,$  and $\,\partial_{IJ} \otimes \partial^{J8} +  \partial^{J8} \otimes \partial_{IJ}  = 0 \,$. Various type IIA/IIB solutions are recovered with a non-trivial dependence on the internal extended space of the form
\begin{equation}
\label{sec_cond_4D_massive}
\begin{array}{lllllll}
i) & \partial_{m 7} \neq 0  &  &    & \hspace{5mm} &\textrm{(type IIA)} & , \\[2mm]
ii) & \partial_{1 7} , ..., \partial_{5 7} \neq 0 & , &  \partial^{6 8} \neq 0  & \hspace{5mm} &\textrm{(type IIB)}  & , \\[2mm]
iii) & \partial_{1 7} , ..., \partial_{4 7} \neq 0 & , &  \partial^{5 8} \,,\,  \partial^{6 8} \neq 0  & \hspace{5mm} &\textrm{(type IIA)}  & , \\[2mm]
iv) & \partial_{1 7} , ..., \partial_{3 7} \neq 0 & , &  \partial^{4 8}, ... \partial^{6 8} \neq 0 & \hspace{5mm}   &\textrm{(type IIB)} & , \\[2mm]
v) & \partial_{1 7} , ..., \partial_{2 7} \neq 0 & , &  \partial^{3 8}, ... \partial^{6 8} \neq 0  & \hspace{5mm}  &\textrm{(type IIA)}& , \\[2mm]
vi) & \partial_{1 7}  \neq 0 & , &  \partial^{2 8}, ... \partial^{6 8} \neq 0  & \hspace{5mm}  &\textrm{(type IIB)} & , \\[2mm]
vii) &  &  &  \partial^{m8} \neq 0  & \hspace{5mm}  &\textrm{(type IIA)} & .
\end{array}
\end{equation}
Note that $vii)$ is actually a type IIA solution embeddable into a \textit{dual M-theory} solution with
\begin{equation}
\label{sec_cond_4D_massive_dual_M}
\begin{array}{lllllll}
viii) & \partial^{I 8} \neq 0  &  & & \hspace{5mm}  &\textrm{(dual M-theory)} & .
\end{array}
\end{equation}
The different cases in (\ref{sec_cond_4D_massive}) are related by a chain of T-duality transformations, in complete analogy with what we found in other XFT's. Starting from $i)$, where the $\XR$ deformation is identified with the Romans mass, one finds
\begin{equation}
\label{T-duality_chain_4D}
\begin{array}{lclclcl}
i) \,\,\, \textrm{massive IIA}  & \,\,\,\,  \overset{T_{67}}{\longrightarrow} \,\,\,\, &
ii) \,\,\, \textrm{IIB with } F_{(1)}  &  \,\,\,\, \overset{T_{57}}{\longrightarrow}  \,\,\,\, &
iii) \,\,\, \textrm{IIA with } F_{(2)}  & \,\,\,\, \overset{T_{47}}{\longrightarrow} \,\,\,\, & \\[2mm]
iv) \,\,\, \textrm{IIB with } F_{(3)} & \overset{T_{37}}{\longrightarrow} &
v) \,\,\, \textrm{IIA with } F_{(4)}  & \overset{T_{27}}{\longrightarrow} & 
vi) \,\,\, \textrm{IIB with } F_{(5)} & \overset{T_{17}}{\longrightarrow} &  \\[2mm]
vii) \,\,\, \textrm{IIA with } F_{(6)} & \leftrightsquigarrow &  \multicolumn{3}{l}{viii) \,\,\, \textrm{dual M-theory with } \,F_{(7)} = \star_{\raisebox{-0.5ex}{\tiny 11D}} \, F_{(4)}}   ,
\end{array}
\end{equation}
where the chain of T-duality transformations $T_{m7}$, with $m=1,...,6$, exchanges the internal coordinates $y^{m7} \leftrightarrow y_{m8}$. The original Romans mass parameter gets consistently mapped into different RR $p$-form fluxes upon T-dualities. In the dual M-theory case, obtained by oxidation of $\textrm{IIA with } F_{(6)}\,$, the $\XR$ deformation corresponds to the Freund--Rubin (FR) parameter \cite{Freund:1980xh}.%
\footnote{Note the difference with the SL(5) XFT discussed before for which a dual M-theory interpretation of the Romans mass was not possible. The reason is that, in the $\,D=7\,$ case, the FR parameter in M-theory maps into a Neveu--Schwarz--Neveu--Schwarz (NSNS) background flux $\,H_{(3)}\,$ in type IIA. The latter is not related to the Romans mass via duality transformations.}

\subsection{Extension to other background fluxes}
\label{sec:SL5_fluxes}

In the previous section we have seen how, starting from a type IIA solution of the section constraint with a non-vanishing Romans mass $\,m_{\rm R}\neq 0\,$,  other type II (or M-theory) background fluxes are obtained upon choosing T-dual solutions (with an extra oxidation). In these dual descriptions, the mass parameter $\,m_{\rm R}\,$ gets consistently mapped into other types of flux parameters which are still compatible with the quadratic constraints \eqref{quadratic constraint}, the section constraint \eqref{section constraint} and the \mbox{$X$-constraint} \eqref{X condition} in XFT. It is therefore natural to wonder whether different types of fluxes can coexist in $\,X\,$ for one choice of solution of the section constraint. This is what we investigate in this section where, for the sake of concreteness, we use again the SL(5) XFT. 
We will select representative M-theory, type IIA and type IIB solutions of the section constraint and find all the $X$ deformations that solve the $\,X$-constraint without imposing further restrictions on the coordinate dependence.

The structure of $X$ deformations in SL(5) XFT parallels that of maximal $D=7$ supergravity \cite{Samtleben:2005bp}. In the latter, deformations are described in terms of an embedding tensor that falls into the $\,\bf{15} + \bf{40}'\,$ irreps of SL(5), thus yielding two pieces $\,Y_{mn}=Y_{(mn)}\,$ and $\,Z^{mn,p}=Z^{[mn],p}\,$ with $\,m=1,...,5\,$ and $\,Z^{[mn,p]}=0\,$. Using these two pieces, one builds an $X$ deformation in XFT of the form
\begin{equation}
\hspace{8mm}
X_{mn \, pq}{}^{rs} = 2 \, X_{mn \, [p}{^{[r}} \, \delta_{q]}^{s]}
\hspace{8mm} \textrm{ with } \hspace{8mm}
X_{mn \, p}{^{r}} = \delta_{[m}^{r} \, Y_{n]p} -2 \, \epsilon_{mnpst} \, Z^{st,r} \ .
\end{equation}

\subsubsection*{Type IIA fluxes in SL(5) XFT}

We start by selecting the type IIA solution of the section constraint in (\ref{sec_cond_7D_massive}) according to which the three internal coordinates are identified with $\,y^{\alpha4}\,$ $\,{(\alpha=1,2,3)}\,$, equivalently $\,\partial_{\alpha 4} \neq 0\,$. An explicit computation shows that the most general $X$ deformation compatible with this solution of the section constraint, as well as with the $X$-constraint, has (independent) non-vanishing components of the form
\begin{equation}
\label{general_eT_7D_IIA}
\tfrac{1}{4} Y_{\alpha 4} = \tfrac{1}{2} \epsilon_{\alpha \beta \gamma } \, Z^{\beta \gamma,5} \equiv  H_{\alpha}
\hspace{2mm} , \hspace{2mm}
Y_{44} \equiv \tfrac{1}{3!}  \epsilon^{\alpha \beta \gamma} \, H_{\alpha \beta \gamma}
\hspace{2mm} , \hspace{2mm}
Z^{5 \alpha , 5} \equiv \tfrac{1}{2} \epsilon^{\alpha \beta \gamma} F_{\beta \gamma} 
\hspace{2mm} , \hspace{2mm}
Z^{45,5} \equiv \tfrac{1}{2} m_{\rm R} \ ,
\end{equation}
thus accounting for $\,3+1+3+1=8\,$ free real parameters. Using the dictionary between the type~IIA fluxes and deformations in Table~\ref{Table:branchings_M/IIA}, the components in (\ref{general_eT_7D_IIA}) are identified with the dilaton ($\,H_{\alpha}\,$), NSNS three-form ($\,H_{\alpha\beta\gamma}\,$) and RR two-form ($\,F_{\alpha\beta}\,$) fluxes, as well as with the Romans mass parameter\footnote{The Romans mass can be dynamically generated in a non-geometric manner (not even locally geometric in the language of ref.~\cite{Aldazabal:2010ef}) by allowing the RR \mbox{one-form} to have a non-trivial dependence on the type IIB coordinates $\,{\tilde{y}_{\alpha} \equiv \frac12 \, \epsilon_{\alpha \beta \gamma} \, y^{\beta\gamma}}\,$ associated with $\,{\tilde{\partial}^{\alpha} \equiv \textbf{3}'_{(1,-1)}}\,$ (see Table~\ref{Table:branchings_IIB}). Using representation theory one finds
\begin{equation*}
m_{\rm R} \equiv \textbf{1}_{(1,-3)} = \textbf{3}'_{(1,-1)} \otimes \textbf{3}_{(0,-2)} \big|_{\textbf{1}}\equiv \tilde{\partial}^{\alpha} A_{\alpha} \ .
\end{equation*}
As discussed in ref.~\cite{Hohm:2011cp} in the context of DFT, the dependence on $\,\tilde{y}_{\alpha}\,$ would violate the section constraint and, in order to recover massive IIA, one would have to explore the non-geometric side of the EFT's where the fields pick up a dependence on  physical and dual coordinates at the same time. We elaborate on this non-geometric approach in Section~\ref{sec:non-geometry}.} $\,m_{\rm R}\,$.

\begin{table}[t!] 
\renewcommand{\arraystretch}{1.25}
\begin{center}
\scalebox{0.88}{
\begin{tabular}{|l|l|l|}
\hline 
$\textrm{SL(5)}$ & $\mathbb{R}^{+}_{1} \times \textrm{SL(4)}$ & $\mathbb{R}^{+}_{1} \times \mathbb{R}^{+}_{2} \times \textrm{SL(3)}$ \\[0.8mm]
\hline 
$\textbf{10}  \,\, \big(\partial_{M}\big)$ & ${\color{blue!70!black}{\textbf{4}_{-\frac{3}{2}}}} \,\, \big(\partial_{i}\big) + \textbf{6}_{1}$ &   $\textbf{1}_{(-\frac32,\frac32)} + {\color{red!70!black}{\textbf{3}_{(-\frac32 ,-\frac12)}}} \,\, \big(\partial_{\alpha}\big) +  \textbf{3}_{(1 ,1)}  +  \textbf{3}'_{(1 ,-1)}$  \\[2mm]
\hline 
$\textbf{24}$  &   ${\color{blue!70!black}{\textbf{4}'_{-\frac{5}{2}}}} \,\, \big(A_{jkl}\big) + \textbf{4}_{\frac{5}{2}} + (\textbf{1}+\textbf{15})_{0}$ &    $\textbf{1}_{(-\frac52,-\frac32)} + {\color{red!70!black}{\textbf{3}'_{(-\frac52,\frac12)}}} \,\, \big(B_{\beta \gamma}\big) + \textbf{1}_{(\frac52,\frac32)} + \textbf{3}_{(\frac52,-\frac12)}$ \\
& &  $+ \, {\color{red!70!black}{\textbf{1}_{(0,0)}}} \,\, \big(\phi\big) + \textbf{8}_{(0,0)} + {\color{red!70!black}{\textbf{3}_{(0,-2)}}} \,\, \big(A_{\beta}\big) +\textbf{3}'_{(0,2)} +  \textbf{1}_{(0,0)} $ \\[2mm]
\hline 
$\textbf{15}\,\, \big(Y_{MN}\big)$  &   ${\color{blue!70!black}{\textbf{1}_{-4} }}  \,\, \big(\partial_{[i}A_{jkl]} \big)+ \textbf{4}_{-\frac{3}{2}} + \textbf{10}_{1}$ & ${\color{red!70!black}{\textbf{1}_{(-4,0)}}}  \,\, \big(\partial_{[\alpha}B_{\beta \gamma]} \big)+ \textbf{1}_{(-\frac32,\frac32)} + {\color{red!70!black}{\textbf{3}_{(-\frac32,-\frac12)}}}  \,\, \big(\partial_{\alpha} \phi \big) $ \\
& & $+ \, \textbf{1}_{(1,3)} +  \textbf{3}_{(1,1)} +  \textbf{6}_{(1,-1)}$    \\[2mm]
\hline 
$\textbf{40}' \,\, \big(Z^{MN,P}\big)$  &  $\textbf{20}'_{-\frac{3}{2}} + \textbf{6}_{1}  + \textbf{10}'_{1}  + \textbf{4}'_{\frac{7}{2}}$ &   $\textbf{8}_{(-\frac32,\frac32)} +  \textbf{6}'_{(-\frac32,-\frac12)} +  {\color{red!70!black}{\textbf{3}'_{(-\frac32,-\frac52)}}}  \,\, \big(\partial_{[\alpha}A_{\beta]} \big)+ {\color{red!70!black}{\textbf{3}_{(-\frac32,-\frac12)} }}  \,\, \big(\partial_{\alpha} \phi \big)$ \\
& &   $+ \, \textbf{3}_{(1,1)} + \textbf{3}'_{(1,-1)} + \underset{m_{\rm R}}{\underline{{\color{red!70!black}{\textbf{1}_{(1,-3)}}}}} + \textbf{3}'_{(1,-1)}  + \textbf{6}'_{(1,1)}$ \\[-3mm]
& &  $+ \,  \textbf{1}_{(\frac72,-\frac32)} + \textbf{3}'_{(\frac72,\frac12)} $ \\[2mm]
\hline 
\end{tabular}}
\caption{Group theory decompositions relevant for the embeddings of M-theory and type IIA into SL(5) XFT. The internal derivatives ($\,\subset \textbf{10}\,$), gauge potentials and dilaton ($\,\subset \textbf{24}\,$) and gauge fluxes ($\,\subset \textbf{15} + \textbf{40}'\,$) are highlighted both in the M-theory (blue) and the natural type~IIA (red) solutions of the section constraint. The Romans mass parameter $\,m_{\rm R}\,$ is singled out. Note that only a linear combination of the two $\,\textbf{3}_{(-\frac32,-\frac12)} \subset \textbf{15} \, , \, \textbf{40}'\,$ is sourced by the dilaton flux $\,\partial_{\alpha}\phi\,$ so that there are $\,1\,$ and $\,8\,$ free real deformation parameters in M-theory and type IIA, respectively.}
\label{Table:branchings_M/IIA} 
\end{center}
\end{table}

The $X$ deformation induced by (\ref{general_eT_7D_IIA}) accounts for all the background \textit{gauge fluxes} that can  thread the three-dimensional internal space. However, this by no means implies that all the parameters can be turned on simultaneously as they still have to obey the quadratic constraints  in (\ref{quadratic constraint}). These take the form of
\begin{equation}
\label{QCmIIA}
m_{\rm R} \, H_{\alpha} = 0
\hspace{8mm} \textrm{ and } \hspace{8mm}
\tfrac{1}{2} \, \epsilon^{\alpha \beta \gamma} \, H_{\alpha} \, F_{\beta \gamma} + \tfrac{1}{4!} \tfrac{1}{3!} \, \epsilon^{\alpha \beta \gamma} \, m_{\rm R} \, H_{\alpha \beta \gamma} = 0 \ ,
\end{equation}
and correspond to the flux-induced tadpole cancellation conditions in absence of O8/D8 and O6/D6 sources, respectively. Solving the quadratic constraints (\ref{QCmIIA}) yields two families of $X$ deformations, equivalently, consistent XFT's. The first one is a six-parameter family of XFT's specified by the two conditions
\begin{align}
&a) \hspace{8mm} 
\epsilon^{\alpha \beta \gamma} \, H_{\alpha} \, F_{\beta \gamma} = 0
\ , \qquad
m_{\rm R} = 0 \ ,
%
\intertext{whereas the second one is a four-parameter family of XFT's specified by the four conditions}
%
&b) \hspace{8mm}
H_{\alpha \beta \gamma}  = 0
\  , \qquad\quad
H_{\alpha} = 0 \ .
\end{align}
As a result, the dilaton flux $\,H_{\alpha}\,$ and the $H_{(3)}$ flux on the one hand,  and the Romans mass parameter $\,m_{\rm R}\,$ on the other cannot be turned on simultaneously.

\begin{table}[t!] 
\renewcommand{\arraystretch}{1.25}
\begin{center}
\scalebox{0.92}{
\begin{tabular}{|l|l|l|}
\hline 
$\textrm{SL(5)}$ &  $\mathbb{R}^{+}_{1} \times \mathbb{R}^{+}_{2} \times \textrm{SL(3)}$ \\[0.8mm]
\hline 
$\textbf{10} \,\,(\partial_{M})$   &   $\textbf{1}_{(-\frac32,\frac32)} + \textbf{3}_{(-\frac32 ,-\frac12)} +  \textbf{3}_{(1 ,1)}  +  {\color{red!70!black}{\textbf{3}'_{(1 ,-1)}}}  \,\, \big(\tilde{\partial}^{\alpha} \equiv \frac{1}{2} \, \epsilon^{\alpha \beta \gamma}\,  \partial_{\beta \gamma}\big)  $  \\[2mm]
\hline 
$\textbf{24}$  &    ${\color{red!70!black}{\textbf{1}_{(-\frac52,-\frac32)}}} \,\, \big(C_{0}\big)  + \textbf{3}'_{(-\frac52,\frac12)} + {\color{red!70!black}{\textbf{1}_{(\frac52,\frac32)}}} \,\, \big(\gamma_{0}\big)  + {\color{red!70!black}{\textbf{3}_{(\frac52,-\frac12)}}} \,\, \big( B^{\beta \gamma}\big)$ \\
&   $+ \, {\color{red!70!black}{\textbf{1}_{(0,0)}}}  \,\, \big(\phi\big) + \textbf{8}_{(0,0)} + {\color{red!70!black}{\textbf{3}_{(0,-2)}}} \,\, \big( C^{\beta \gamma}\big)+ \textbf{3}'_{(0,2)} +  \textbf{1}_{(0,0)} $ \\[2mm]
\hline 
$\textbf{15}\,\,(Y_{MN})$ & $\textbf{1}_{(-4,0)} + \textbf{1}_{(-\frac32,\frac32)} + \textbf{3}_{(-\frac32,-\frac12)} + \textbf{1}_{(1,3)} +  \textbf{3}_{(1,1)} +  \textbf{6}_{(1,-1)}$    \\[2mm]
\hline 
$\textbf{40}' \,\,(Z^{MN,P})$   &   $\textbf{8}_{(-\frac32,\frac32)} +  \textbf{6}'_{(-\frac32,-\frac12)} +  {\color{red!70!black}{\textbf{3}'_{(-\frac32,-\frac52)}}} \,\, \big(\tilde{\partial}^{\alpha} C_{0}\big)+ \textbf{3}_{(-\frac32,-\frac12)}$ \\
&    $+ \, \textbf{3}_{(1,1)} + {\color{red!70!black}{\textbf{3}'_{(1,-1)}}} \,\, \big(\tilde{\partial}^{\alpha} \phi \big) + {\color{red!70!black}{\textbf{1}_{(1,-3)}}}\,\, \big(\tilde{\partial}^{[\alpha} C^{\beta \gamma]} \big)  + {\color{red!70!black}{\textbf{3}'_{(1,-1)}}}\,\, \big(\tilde{\partial}^{\alpha} \phi\big)  + \textbf{6}'_{(1,1)}$ \\
&   $+ \,  {\color{red!70!black}{\textbf{1}_{(\frac72,-\frac32)}}} \,\, \big(\tilde{\partial}^{[\alpha} B^{\beta \gamma]} \big)+ {\color{red!70!black}{\textbf{3}'_{(\frac72,\frac12)}}} \,\, \big(\tilde{\partial}^{\alpha} \gamma_{0}\big) $ \\[5mm]
\hline 
\end{tabular}}
\caption{Group theory decompositions relevant for the embedding of type IIB into SL(5) XFT. The purely internal derivatives ($\,\subset \textbf{10}\,$), gauge potentials and scalars ($\,\subset \textbf{24}\,$) and gauge fluxes ($\,\subset \textbf{40}'\,$) are highlighted (red). Note that only a linear combination of the two $\,\textbf{3}'_{(1,-1)} \subset \textbf{40}'\,$ is sourced by the dilaton flux $\,\tilde{\partial}^{\alpha}\phi\,$ so that there are $\,11\,$ free real deformation parameters in type IIB. The $\,\mathbb{R}^{+}_{\textrm{S}} \in \textrm{SL}(2)\,$ charge of the type~IIB theory (S-duality) is given by $\,q_{\textrm{S}}=q_{\mathbb{R}^{+}_1}+q_{\mathbb{R}^{+}_2}\,$.}
\label{Table:branchings_IIB} 
\end{center}
\end{table}

\subsubsection*{M-theory fluxes in SL(5) XFT}

The same analysis can be performed for the M-theory extension of the type IIA solution in (\ref{sec_cond_7D_massive}). In this case, the four internal coordinates $\,y^{i}\,$ of the eleven-dimensional supergravity are identified with $\,y^{\alpha 4}\,$ and $\,y^{45}\,$, the latter being the M-theory coordinate. The most general $X$ deformation compatible with the $X$-constraint has a unique non-vanishing component given by 
\begin{equation}
Y_{44} \equiv f_{\textrm{FR}} \ ,
\end{equation}
and is identified (see Table~\ref{Table:branchings_M/IIA}) with the Freund--Rubin parameter \cite{Freund:1980xh}. This parameter corresponds to a purely internal background for the field strength of the three-form potential of eleven-dimensional supergravity and is compatible with the quadratic constraint in (\ref{quadratic constraint}). Therefore, there is a one-parameter family of XFT's that describes such eleven-dimensional backgrounds with an $\,f_{\textrm{FR}}\,$ flux.

\subsubsection*{Type IIB fluxes in SL(5) XFT}

Lastly there is the type IIB case in (\ref{sec_cond_7D_massive}) where the three internal coordinates are identified with $\,\tilde{y}_{\alpha} \equiv \frac12 \, \epsilon_{\alpha \beta \gamma} \, y^{\beta \gamma}\,$, equivalently, $\,\tilde{\partial}^{\alpha}\neq 0\,$. The most general $X$ deformation satisfying the $X$-constraint is compatible with the SL(2) symmetry (S-duality) of the IIB theory and has (independent) non-vanishing components of the form
\begin{equation}
\label{general_eT_7D_IIB}
\begin{array}{rcc}
\textrm{SL(2)-doublet :} & & Z^{45,5} \equiv \tfrac{1}{3!} \, \epsilon_{\alpha \beta \gamma} \, F^{\alpha \beta\gamma}
\hspace{4mm} , \hspace{4mm}
Z^{45,4} \equiv \tfrac{1}{3!} \, \epsilon_{\alpha \beta \gamma} \,  H^{\alpha \beta\gamma} \ ,  \\[2mm]
\textrm{SL(2)-triplet :} & & Z^{\alpha5,5} \equiv  F^{\alpha}
\hspace{4mm} , \hspace{4mm}
Z^{\alpha 5,4} = Z^{\alpha 4,5}  \equiv  H^{\alpha}
\hspace{4mm} , \hspace{4mm}
Z^{\alpha 4,4} \equiv  \hat{F}^{\alpha}  \ ,

\end{array}
\end{equation}
accounting for $\,2 \times 1 \, + \, 3 \times 3 \,=\, 11 \,$ free real parameters. Using the dictionary between type~IIB fluxes and deformations in Table~\ref{Table:branchings_IIB}, one identifies an SL(2)-doublet of RR ($\,F^{\alpha\beta\gamma}\,$) and NSNS  ($\,H^{\alpha\beta \gamma}\,$) three-form fluxes\footnote{See also ref.~\cite{Blair:2014zba} for a discussion on generalised fluxes in SL(5) EFT.}. In addition, there is also an SL(2)-triplet of one-form deformations\footnote{In the `gauge-unfixed' approach of ref.~\cite{Aldazabal:2010ef}, one may consider an additional scalar $\,\gamma_{0}\equiv \textbf{1}_{(\frac52,\frac32)}\,$. Using again representation theory, one finds
\begin{equation*}
\hat{F}^{\alpha} \,\equiv\,  \textbf{3}'_{(\frac72,\frac12)} \,=\, \textbf{3}'_{(1,-1)} \otimes \textbf{1}_{(\frac52,\frac32)} \,\, \oplus \,\, \textbf{3}_{(1,1)} \otimes \textbf{3}_{(\frac52,-\frac12)}  \big|_{\textbf{3}'} \, \equiv \, \tilde{\partial}^{\alpha} \gamma_{0}  \,\oplus \, \hat{\partial}_{\beta} B^{\beta \alpha} \ ,
\end{equation*}
which includes two different types of contributions. In the language of ref.~\cite{Aldazabal:2010ef}, the first term in the r.h.s. corresponds to a locally geometric way of generating $\,\hat{F}^{\alpha}\,$ by turning on a flux for the spurious scalar $\,\gamma_{0}\,$. The second term  is generated when the NSNS two-form potential depends on the dual coordinates $\,\hat{y}^{\alpha}\,$, namely  $\,\hat{\partial}_{\beta} B^{\beta \alpha} \neq 0\,$, with $\,\hat{\partial}_{\alpha} \equiv  \textbf{3}_{(1,1)}\,$. Note that these are \textit{not} the type IIA coordinates (see Table~\ref{Table:branchings_M/IIA}). This case is analogous to what happens in type IIA for the Romans mass and produces an $\,\hat{F}^{\alpha}\,$ flux which is not even locally geometric, thus violating the section constraint in EFT.} $\,(F^{\alpha},H^{\alpha},\hat{F}^{\alpha})\,$. The latter account for an internal dependence of the type IIB axion-dilaton and can be dualised into nine-form fluxes for the SL(2)-triplet of RR eight-form potentials of the IIB theory \cite{Meessen:1998qm,Dall'Agata:1998va}.

The computation of the quadratic constraints in (\ref{quadratic constraint}) for the type IIB fluxes in (\ref{general_eT_7D_IIB}) produces the set of relations
\begin{equation}
\label{tadpoles7}
\epsilon_{\alpha \beta \gamma} \, F^{\beta} \, H^{\gamma} = 0
\hspace{5mm} , \hspace{5mm} 
\epsilon_{\alpha \beta \gamma} \, F^{\beta} \, \hat{F}^{\gamma} = 0
\hspace{5mm} \textrm{ and } \hspace{5mm} 
\epsilon_{\alpha \beta \gamma} \, \hat{F}^{\beta} \, H^{\gamma} = 0 \ ,
\end{equation}
which corresponds to flux-induced tadpole cancellation conditions for an SL(2)-triplet of \mbox{7-branes} (and related orientifold planes). Again such objects must be absent in order to preserve maximal supersymmetry. Note that (\ref{tadpoles7}) is SL(2)-covariant and can be rewritten as $\,{\mathcal{H}_{[A} \wedge \mathcal{H}_{B]}= 0}\,$ with $\,A=1,2,3\,$ and $\,\mathcal{H}_{A}=(F,H,\hat{F})\,$. Solving (\ref{tadpoles7}) yields a seven-parameter family of XFT's that describes such ten-dimensional type IIB backgrounds. 

Let us close the section commenting on the number of deformation parameters permitted in other XFT's. For a given $\,D \ge 4\,$, the most general $X$ deformation compatible with the section constraint \eqref{section constraint} and \mbox{$X$-constraint} \eqref{X condition} includes: $i)$ the Freund--Rubin parameter in M-theory (only for $\,D=7,4\,$) \, $ii)$ the Romans mass $\,m_{\rm R}\,$ (any $D$) as well as dilaton (any $D$) and standard $p$-form gauge fluxes (whenever permitted by $D$) in type IIA \, $iii)$ the SL(2)-triplet of one-form deformations (any $D$) as well as standard $p$-form gauge fluxes (whenever permitted by $D$) in type IIB. In order to specify a consistent XFT, the resulting $X$ deformation must still be supplemented with the quadratic constraint (\ref{quadratic constraint}). This can be translated into tadpole cancellation conditions requiring the absence of sources of supersymmetry breaking.  A last remark concerns the incompatibility of the $X$-constraint with the presence of a metric flux $\,\omega\,$ of the Scherk--Schwarz (SS) type \cite{Scherk:1979zr}. 
Suppose that it was possible to introduce $\,\omega\,$ in XFT while allowing fields to depend \emph{arbitrarily} on ten or eleven physical coordinates after solution of the $\,X$-constraint.
The resulting $\,X$-deformation would modify the action of \emph{ordinary} internal diffeomorphisms rather than that of $p$-form gauge transformations, which is not possible.
Consistently, we find that the $\,X$-constraint actually excludes metric fluxes.

\section{Dynamics of $\textrm{E}_{7(7)}$ XFT}
\label{sec:E7-XFT}

In this section we illustrate the generic features of the deformations introduced in Section~\ref{sec:Gen_diff_X} by constructing explicitly the gauge invariant E$_{7(7)}$ XFT. While the field content of the theory remains identical to the one of E$_{7(7)}$ EFT, changes occur at the level of the tensor hierarchy and in the action due to the presence of the $X$ deformation. We refer to Section \ref{subsec:SC} for a detailed discussion of the section constraint of the E$_{7(7)}$ XFT. We present below some specifics of the deformed E$_{7(7)}$ generalised diffeomorphisms, followed by the tensor hierarchy and the full bosonic action. The latter consistently reduces to the action of $\,D=4\,$ gauged maximal supergravity when all fields are taken independent of the \textbf{56} exceptional coordinates $\,y^{M}\,$, and to the one of the E$_{7(7)}$ EFT when the $X$ deformation is turned off. 
Finally, when fixing $\,X\,$ to \eqref{X_4D_Romans} and choosing an appropriate solution of the section and $X$-constraints (see \eqref{sec_cond_4D_massive}), one recovers the bosonic sector of massive type IIA supergravity in a $4+6$ dimensional split. The results of this section are in parallel with those of ref.~\cite{Hohm:2013uia} to which we refer for an in-depth discussion of the E$_{7(7)}$ EFT dynamics.

\subsection{Modified Lie derivative and trivial parameters}

The expression of the E$_{7(7)}$ invariant  $Y$-tensor is given in \eqref{Y-tensor_4D}.
Both the \Eseven generators $[t_{\alpha}]_{M}{}^N$ and the symplectic form $\,\Omega_{MN}\,$ are  invariant under the deformed generalised Lie derivative \eqref{eq:def}.
For E$_{7(7)}$ the distinguished weight to be introduced in \eqref{def_Lie_1} is $\,\omega=\tfrac12\,$. 
The section constraint decomposes into two irreducible pieces in the ${\bf1}+{\bf133}$ irreps:
\begin{formula}
\Omega^{MN}\partial_M\otimes\partial_N=0\,,\qquad
[t_\alpha]^{MN}\partial_M\otimes\partial_N=0\,.
\end{formula}
We will use the shorthand notation $(\bbP_{{\bf1}+{\bf133}})^{MN}\partial_M\otimes\partial_N=0$ to reflect these two constraints.

As explained previously, the $X$ deformation satisfies the same linear and quadratic constraints as the embedding tensor in gauged maximal supergravity \cite{deWit:2007mt}.
The linear constraints in $D=4$ read
\begin{equation}
\label{eq:lin}
X_{NM}{}^M=X_{MN}{}^M=0\,,\qquad
X_{(MNP)}=0\,,
\end{equation} 
and restrict $X$ to belong to the $\bf912$ representation.
Consequently, the quadratic constraint \eqref{quadratic constraint} can be rephrased as
\begin{formula}
\label{eq:locality}
\Omega^{MN}\,X_M \otimes X_N = 0\,.
\end{formula}
The deformed EFT requires to impose the section- and the  $X$-constraint \eqref{section constraint}.
In $D=4$ the latter can be decomposed into the ${\bf 133}+{\bf 1539}$ irreps, corresponding to $`X_(MN)^P \partial_P $ and $`X_[MN]^P \partial_P $ respectively.
Using representation theory it is possible to find other equivalent ways to express these constraints.
Two such expressions are particularly useful
\begin{formula}
\label{4d equiv X constraints}
\Omega^{MN} \Theta_M{}^\alpha \partial_N = 0\,,\qquad	
\Theta_M{}^{(\alpha} [t^{\beta)}]^{MN} \partial_N = 0\,,
\end{formula}
and correspond to the ${\bf 133}$ and ${\bf 1539}$, respectively.

The construction of the \Eseven XFT tensor hierarchy relies on the form of certain trivial parameters appearing in the symmetric $X$-bracket \eqref{eq:symbracket}.
Specifically, for two arbitrary generalised vectors of weight $\omega$ we have
\begin{formula}
\label{eq:trivparaex}
\big\{U,V\big\}^M_X =&\, - 6 \, [t_\alpha]^{MN}[t^\alpha]_{PQ}\,\partial_N\big[U^P V^Q\big]-U^N V^P X_{(NP)}{}^M \\
&\, - \frac14 \, \Omega^{MN} \, \Omega_{PQ}\big[V^Q\,\partial_N U^P+U^Q\,\partial_NV^P\big] \ .
\end{formula}
Both lines of \eqref{eq:trivparaex} are trivial parameters provided all fields satisfy the section constraint and the symmetric part of the $X$-constraint \eqref{X condition}. This ensures that the Jacobi identity for $\,\widetilde\bbL\,$ is satisfied.
More generally, the following generic parameters do not generate generalised diffeomorphisms:
\begin{align}
&\Lambda^M=[t^\alpha]^{MN}\partial_{N}\chi_\alpha +\frac16 \, Z^{M,\alpha} \, \chi_\alpha\,,\label{eq:trivpara1}\\
&\Lambda^M=\Omega^{MN}\chi_N \ ,\label{eq:trivpara2}
\end{align}
for arbitrary $\,\chi_\alpha\,$. The intertwining tensor $Z^{M,\alpha}$ is constructed from $`X_MN^P $ making use of the linear constraint:
\begin{equation}
Z^{M,\alpha} = - X_{PQ}{}^M \, [t^\alpha]^{PQ} = -\frac12 \Omega^{MN}\Theta_N{}^\alpha\ .
\end{equation}
Similarly to the EFT case, $\,\chi_M\,$ is covariantly constrained in the sense that it must itself satisfy the section constraints 
\begin{equation}
\label{eq:covconstraint}
\big(\mathbb{P}_{\mathbf{1}+\mathbf{133}}\big)^{MN}\chi_M\partial_N\,=\,
0\,=\,\big(\mathbb{P}_{\mathbf{1}+\mathbf{133}}\big)^{MN}\chi_M\chi_N\,,
\end{equation}
where $\,\mathbb{P}_{\mathbf{1}+\mathbf{133}}\,$ denotes the projector onto the $\,\mathbf{1}\oplus\mathbf{133}\,$ representation of the $\,\mathbf{56}\otimes\mathbf{56}\,$. In XFT, the field $\chi_M$ is further covariantly constrained by 
\begin{equation}
\label{eq:covconstraint2}
X_{(MN)}{}^P\chi_P=0\,,
\end{equation}
or equivalently by $\,\Omega^{MN}\Theta_{M}{}^\alpha \,\chi_N=0\,$. The importance of the covariantly constrained parameters \eqref{eq:trivpara2} will become apparent when constructing the tensor hierarchy.

\subsection{Yang-Mills sector and tensor hierarchy}

Analogously to EFT, we introduce an external derivative which is covariant under modified internal generalised diffeomorphisms
\begin{equation}
\mathcal{D}_{\mu} \equiv \partial_{\mu} - \widetilde{\mathbb{L}}_{A_{\mu}} \ .
\end{equation}
Covariance determines the variation of $\,A_\mu{}^M\,$ to be 
\begin{equation}
\label{deltaAM_1_X}
\delta_\Lambda A_\mu{}^M=\mathcal{D}_\mu \Lambda^M 
\simeq
\partial_\mu\Lambda^M + \widetilde{\mathbb{L}}_\Lambda A_\mu{}^M \ ,
\end{equation}
where the equivalence holds up to the addition of a trivial gauge parameter, to be reabsorbed in other gauge transformations higher up in the tensor hierarchy.
This is completely in line with the situation for the undeformed EFT. 

Following the construction of the tensor hierarchy in the original EFT's, we first define the field strength for the vector fields $\,A_\mu{}^M\,$ as
\begin{equation}
F_{\mu\nu}{}^M = 2 \, \partial_{[\mu}A_{\nu]}{}^M-\big[A_{\mu},A_{\nu}\big]_X^M \ .
\end{equation}
Since the Jacobiator of the $\,X$-bracket does not vanish, the above expression does not transform covariantly under generalised diffeomorphisms.
The procedure to restore gauge covariance is analogous to those of gauged supergravity and EFT.
In fact, it turns out to be a superposition of the two cases.
We define a modified field strength by introducing the two-form fields $\,B_{\mu\nu\,\alpha}\,$ and $\,B_{\mu\nu\,M}\,$ in the form of the two
trivial parameters \eqref{eq:trivpara1} and \eqref{eq:trivpara2}
\begin{equation}
\label{eq:modfieldstren}
\mathcal{F}_{\mu\nu}{}^M=F_{\mu\nu}{}^M - 12 \, [t^\alpha]^{MN}\partial_NB_{\mu\nu\,\alpha} - 2 \, Z^{M,\alpha}B_{\mu\nu\,\alpha}-\frac12  \, \Omega^{MN}B_{\mu\nu\,N}\ ,
\end{equation}
where $\,B_{\mu\nu\,K}\,$ is a covariantly constrained field as in \eqref{eq:covconstraint} and \eqref{eq:covconstraint2}. Note that this construction only deviates from EFT by the term proportional to $\,Z^{M,\alpha}\,$, which is precisely the one needed to make contact with gauged supergravities when all the fields are taken to be $y^M$-independent. 
It is easy to verify that, since $\,F_{\mu\nu}{}^M\,$ only differs from $\,\mathcal{F}_{\mu\nu}{}^M\,$ by a trivial parameter, we have
\begin{equation}
\label{eq:Xricci}
\big[\mathcal{D}_{\mu},\mathcal{D}_\nu\big]=-2\,\widetilde{\mathbb{L}}_{\partial_{[\mu}A_{\nu]}}+2\,\widetilde{\mathbb{L}}_{A_{[\mu}} \widetilde{\mathbb{L}}_{A_{\nu]}} = - \widetilde{\mathbb{L}}_{F_{\mu\nu}}=-\widetilde{\mathbb{L}}_{\mathcal{F}_{\mu\nu}} \ .
\end{equation}
Using the explicit expression for the symmetric $X$-bracket \eqref{eq:trivparaex}, the general variation of the modified field strength \eqref{eq:modfieldstren} now reads
\begin{equation}
\label{eq:modfieldstrenvar}
\delta \mathcal{F}_{\mu\nu}{}^M=2 \, \mathcal{D}_{[\mu}\delta A_{\nu]}{}^M-12 \, [t^\alpha]^{MN}\partial_{N}\Delta B_{\mu\nu\,\alpha}-2\,Z^{M,\alpha}\Delta B_{\mu\nu\,\alpha}-\frac12 \, \Omega^{MN} \, \Delta B_{\mu\nu\,N}\,,
\end{equation}
where, as in EFT, we have defined
\begin{formula}
\Delta B_{\mu\nu\,\alpha} &=\, \delta B_{\mu\nu\,\alpha}+[t_\alpha]_{NP} A_{[\mu}{}^N\delta A_{\nu]}{}^P \ ,\\
\Delta B_{\mu\nu\, N} &=\, \delta B_{\mu\nu\,N}+\Omega_{PQ}\big[A_{[\mu}{}^Q\partial_N\delta A_{\nu]}{}^P+\partial_N A_{[\mu}{}^P\delta A_{\nu]}{}^Q\big] \ .
\end{formula}
We define the vector gauge variations of the two-forms as follows\footnote{It will be convenient for compatibility with \cite{Hohm:2013uia} to take $\delta_\Lambda A_\mu{}^M=\mathcal{D}_\mu \Lambda^M$ as the variation for the vector fields under generalised diffeomorphisms (cfr. the discussion below \eqref{deltaAM_1_X}).}:
\begin{formula}
\Delta_\Lambda  B_{\mu\nu\,\alpha}=&\,[t_\alpha]_{NP} \Lambda^N\mathcal{F}_{\mu\nu}{}^P\,,\\
\Delta_\Lambda B_{\mu\nu\, N}=&\,\Omega_{PQ}\big[\Lambda^Q\partial_N\mathcal{F}_{\mu\nu}{}^{P}+\mathcal{F}_{\mu\nu}{}^Q\partial_{N}\Lambda^P\big]\,.
\end{formula}
Substituting the above variations back in \eqref{eq:modfieldstrenvar} and making use of \eqref{eq:trivparaex} and \eqref{eq:Xricci} yields
\begin{equation}
\label{eq:Finv}
\delta_\Lambda \mathcal{F}_{\mu\nu}{}^M = \big[\mathcal{D}_{\mu},\mathcal{D}_{\nu}\big]\Lambda^M+2\big\{\Lambda,\mathcal{F}_{\mu\nu}\big\}_X^M = \widetilde{\mathbb{L}}_{\Lambda}\mathcal{F}_{\mu\nu}{}^M\,,
\end{equation}
which shows that $\,\mathcal{F}_{\mu\nu}\,$ transforms covariantly.

On top of the generalised diffeomorphisms (\textit{i.e.} vector gauge transformations), the field strength \eqref{eq:modfieldstren} is invariant under tensor gauge transformations associated with the two-forms
\begin{formula}
\delta_{\Xi}A_{\mu}{}^M=&\,12 [t^\alpha]^{MN}\partial_N \Xi_{\mu\,\alpha}+2\, Z^{M,\alpha}\,\Xi_{\mu\,\alpha}+\frac12 \, \Omega^{MN}\Xi_{\mu\,N}\,,\\
\Delta_{\Xi}B_{\mu\nu\,\alpha}=&\,2\,\mathcal{D}_{[\mu}\Xi_{\nu]\alpha}\,,\\
\Delta_{\Xi}B_{\mu\nu\,M}=&\,2\,\mathcal{D}_{[\mu}\Xi_{\nu]M}+48\,[t^\alpha]_L{}^K(\partial_K\partial_MA_{[\mu}{}^L)\Xi_{\nu]\alpha}+4\,\Theta_P{}^\alpha\,\partial_M A_{[\mu}{}^P\, \Xi_{\nu]\alpha} \,,
\end{formula}
where the tensor gauge parameters $\,\Xi_{\mu\,\alpha}\,$ and $\,\Xi_{\mu\,M}\,$ carry weight $1$ and $\tfrac12$,  respectively.  
For an arbitrary generalised vector $W_{\alpha}$ in the adjoint of E$_{7(7)}$ with weight $\lambda^\prime$, the deformed generalised Lie derivative acts as follows:  
\begin{equation}
\widetilde{\mathbb{L}}_{\Lambda} W_\alpha =\,\Lambda^R\partial_R W_\alpha -12\,f_{\gamma\alpha}{}^\beta[t^\gamma]_L{}^K\,\partial_K\Lambda^L W_\beta+\lambda^\prime\partial_R\Lambda^R W_\alpha-\Lambda^N \Theta_N{}^\gamma f_{\gamma\alpha}{}^\beta W_\beta \ ,
\end{equation}
where we have used the definition \eqref{eq:def} and the relation between the generators in the adjoint and the structure constants $[t_{\gamma}]_\alpha{}^\beta=-f_{\gamma\alpha}{}^\beta$. 
In order to verify the invariance of the field strength under tensor gauge transformations, it is necessary to study the following expression in terms of a covariant object $W_\alpha\,$:
\begin{equation}
T^M\equiv[t^\alpha]^{MN}\partial_N W_\alpha +\frac16 \, Z^{M,\alpha} \, W_\alpha \ .
\end{equation} 
Under generalised diffeomorphisms, it transforms as 
\begin{formula}
\label{eq:comb}
\delta_\Lambda T^M=\,&\,\widetilde{\mathbb{L}}_\Lambda T^M+\Omega^{MN} \, \big( \, [t^\alpha]_L{}^{K}W_\alpha\partial_N\partial_K\Lambda^K+\tfrac{1}{12} \, \Theta_P{}^\alpha W_\alpha \partial_N\Lambda^P \, \big) \\
&+(\lambda^\prime-1)[t^\alpha]^{MN}W_\alpha\partial_N\partial_K\Lambda^K \ .
\end{formula}
where $\,T^M\,$ carries weight $\,\lambda(T^M)=(\lambda^\prime-\tfrac12)\,$. As in ref.~\cite{Hohm:2013uia}, in order to cancel the non-covariant terms in the first line, a compensating field $W_M$ subject to the covariant constraints \eqref{eq:covconstraint},\eqref{eq:covconstraint2} is introduced such that the combination 
\begin{equation}
\label{eq:combcov}
\hat{T}^M\equiv T^M+\frac{1}{24} \, \Omega^{MN} \, W_N \ ,
\end{equation} 
transforms covariantly with  $\lambda(\hat T^M)=\tfrac12$ provided that $\lambda^\prime=1$.  
This is ensured only if the compensating field transforms under generalised diffeomorphisms as 
\begin{equation}
\label{eq:trans}
\delta_\Lambda W_M = \widetilde{\mathbb{L}}_\Lambda W_M - 24 \, [t^\alpha]_L{}^KW_\alpha\partial_M\partial_K\Lambda^L - 2 \, \Theta_P{}^\alpha\, W_\alpha \partial_M\Lambda^P \ ,
\end{equation}
where $\,\lambda(W_M)=\tfrac12\,$. Note that \eqref{eq:trans} preserves the covariant constraints \eqref{eq:covconstraint} and \eqref{eq:covconstraint2} by virtue of the section constraint (\ref{section constraint}) and the $X$-constraint (\ref{X condition}). With the observation that structures of the form \eqref{eq:combcov} transform covariantly, it becomes straightforward to verify the invariance of the field strength under both tensor gauge transformations.

The field strengths $\,H_{\mu\nu\rho\,\alpha}\,$ and $\,H_{\mu\nu\rho\,M}\,$ associated to the two-forms are defined through the Bianchi identity
\begin{equation}
\label{eq:Bianchi}
3\,\mathcal{D}_{[\mu}\mathcal{F}_{\nu\rho]}{}^M = -12 \, [t^\alpha]^{MN}\partial_N H_{\mu\nu\rho\,\alpha}-2 \,Z^{M,\alpha}H_{\mu\nu\rho\,\alpha} - \frac12 \, \Omega^{MN} \, H_{\mu\nu\rho\,N}\,,
\end{equation}
up to terms that get projected out under $\,6 \, [t^\alpha]^{MN}\partial_N+Z^{M,\alpha}\,$. The field strength $\,H_{\mu\nu\rho\,M}\,$ is again covariantly constrained as in \eqref{eq:covconstraint} and \eqref{eq:covconstraint2} and transforms according to \eqref{eq:trans} under generalised diffeomorphisms.

\subsection{Bosonic action}

In analogy with ref.~\cite{Hohm:2013uia},
the full dynamics of the theory can be encoded into an \Eseven covariant (pseudo-)action supplemented by a first-order duality equation for the \textbf{56} gauge fields $A_{\mu}{}^M\,$ 
\begin{equation}
\label{eq:dualityeq}
\mathcal{F}_{\mu\nu}{}^M = - \tfrac12 \, e \, \varepsilon_{\mu\nu\rho\sigma} \, \Omega^{MN} \, \mathcal{M}_{NK}\mathcal{F}^{\rho\sigma K} \ ,
\end{equation}
where $\,e\,$ denotes the determinant of the vierbein and $\,\mathcal{M}_{MN}\,$ is the scalar matrix parameterising the coset space $\,\textrm{E}_{7(7)}/\textrm{SU}(8)\,$. This ensures that only half of the vectors are independent.

The field equations can be conveniently derived by varying the following gauge invariant pseudo-action, and subsequently imposing \eqref{eq:dualityeq}:
\begin{formula}
\label{eq:action}
S_{\text{\tiny{XFT}}} =\, \displaystyle\int d^4 x\,d^{56} y\,e\,\big[\,\hat{R}&(X)\,+\,\frac{1}{48} \, g^{\mu\nu}\,\mathcal{D}_\mu\mathcal{M}^{MN}\,\mathcal{D}_{\nu}\mathcal{M}_{MN} \\
&\,-\frac18 \, \mathcal{M}_{MN}\,\mathcal{F}^{\mu\nu M}\mathcal{F}_{\mu\nu}{}^N+\,e^{-1}\,\mathcal{L}_{\text{top}}(X)-V_{\text{\tiny{XFT}}}(\mathcal{M},g,X)\,\big] \, .
\end{formula}
%
For the purpose of this paper we shall always assume that integration over the internal space is actually performed only on the physical coordinates after choosing a solution of the section constraint, so that global integration over the internal manifold is well defined.
While the general form of the action matches the one of EFT, the differences with the latter lie in the expressions of the field strengths, the covariant derivatives and the `scalar potential' which explicitly depend on the $X$ deformation. As in EFT, the XFT action is uniquely determined by requiring gauge invariance under the bosonic symmetries. More specifically, each term in \eqref{eq:action} is invariant under internal generalised diffeomorphisms while the relative coefficients are fixed by external diffeomorphisms.

In what follows we discuss the invariance of the different terms under vector (\textit{i.e.} generalised diffeomorphisms) and tensor gauge transformations. 
In the forthcoming computations we will consistently drop all the vector gauge transformations of scalar density of weight 1. Indeed, these take the form of boundary terms in the extended space.

\paragraph{The kinetic terms:} The first term in the action is the Einstein-Hilbert term. As in EFT, it is built from a modified Riemann tensor
\begin{equation}
\label{eq:riemannim}
\hat{R}_{\mu\nu}{}^{ab}(X)=\,R_{\mu\nu}{}^{ab}[\omega]+\mathcal{F}_{\mu\nu}{}^M \,e^{a\rho}\,\partial_M e_\rho{}^b \ ,
\end{equation}
where the curvature of the four dimensional spin connection $\,\omega_{\mu}{}^{ab}\,$ reads
\begin{equation}
R_{\mu\nu}{}^{ab}[\omega]=\,2\,\mathcal{D}_{[\mu}\omega_{\nu]}{}^{ab}-2\,\omega_{[\mu}{}^{ac}\,\omega_{\nu]c}{}^b\ .
\end{equation}
The second term in \eqref{eq:riemannim} has been added in order for the modified Riemann tensor to transform covariantly under the four dimensional local Lorentz transformations acting on the spin connection as 
$\delta_{\lambda}\omega_{\mu}{}^{ab}=-\mathcal{D}_{\mu}\lambda^{ab}$.
The spin connection can in turn be expressed via Cartan's (covariantised) first structure equation in terms of the vierbein $e_{\mu}{}^a$ which is an $\,\textrm{E}_{7(7)}\,$ scalar of weight $\frac12$. Consequently, the spin connection and the Riemann tensor both carry weight 0. Furthermore, using the section constraint and the $X$-constraint, it is straightforward to show that the internal derivative of an $\textrm{E}_{7(7)}$ scalar $\,S\,$ of weight $\,\lambda(S)\,$ transforms under vector gauge transformations as
\begin{equation}
\label{eq:varscal}
\delta_\Lambda (\partial_M S) = \widetilde{\mathbb{L}}_\Lambda (\partial_M S)+\lambda(S)\, S \,\partial_M\partial_N \Lambda^N\,,\;\;\;\;\;\;\text{with weight}\;\;\;\lambda(\partial_M S)=\lambda(S)-\tfrac12 \ .
\end{equation} 
Hence, the modified Riemann tensor does not transform covariantly due to the second term in \eqref{eq:riemannim}. The non-covariant part of the variation vanishes when contracted with vierbeine and therefore, the modified Ricci scalar $\,\hat{R}(X)\,$ is a scalar of weight 0. This proves the invariance of the XFT Einstein-Hilbert term under gauge transformations.

The second and third term in \eqref{eq:action} are respectively the kinetic terms for the scalars and the vector fields. They only differ from the ones in EFT by the implicit presence of the $X$ deformation. The scalar matrix $\,\mathcal{M}_{MN}\,$ is a tensor of weight $\,0\,$ while $\,\mathcal{F}_{\mu\nu}{}^N\,$ carries weight $\,\tfrac12\,$. Using \eqref{eq:Finv} and $\,\delta_\Xi \mathcal{F}_{\mu\nu}{}^M=0\,$, it is clear that both terms are invariant under vector and tensor gauge transformations.

\paragraph{The topological term:} Following ref.~\cite{Hohm:2013uia}, we present the topological term as a surface term in five spacetime dimensions
\begin{formula}
S_{\text{top}}(X)&=-\frac{1}{24}\int_{\Sigma^5}d^5x\int d^{56} y \,\varepsilon^{\mu\nu\rho\sigma\tau}\mathcal{F}_{\mu\nu}{}^M\,\mathcal{D}_{\rho}\mathcal{F}_{\sigma\tau\,M}\\
&\equiv\int_{\partial\Sigma^5}d^5 x\int d^{56} y \,\mathcal{L}_{\text{top}}(X)\,,
\end{formula}
where once again the difference with EFT lies in the definition of the field strength and the covariant derivative. Although this term is manifestly gauge invariant, its general variation is needed to derive the field equations for the vectors and two-forms
\begin{formula}
\label{eq:vartop}
\delta \mathcal{L}_{\text{top}}=-&\frac14 \, \varepsilon^{\mu\nu\rho\sigma}\big[\delta A_{\mu}{}^M\,\mathcal{D}_{\nu}\mathcal{F}_{\rho\sigma\,M}\\[1ex]
&\,\;+\mathcal{F}_{\mu\nu\,M}\big(6 \, [t^\alpha]^{MN}\partial_{N}\Delta B_{\rho\sigma\,\alpha}+Z^{M,\alpha}\Delta B_{\rho\sigma\,\alpha}+\tfrac14 \, \Omega^{MN}\Delta B_{\rho\sigma\,N}\big)\big] \, .
\end{formula}
This requires to use the Bianchi identity \eqref{eq:Bianchi} and the fact that for any three vectors of weight $\tfrac12$ the following identity holds
\begin{equation}
\label{eq:EFT cyclic identity}
\Omega_{MN}\,U^{M}\big\{V,W\big\}_X^N+\text{cyclic}\, =
12\,[t_{\alpha}]_{(M}{}^Q [t^\alpha]_{NP)}\,\partial_Q(U^MV^NW^P)\,.
\end{equation}
The $X$-dependent part of the l.h.s. vanishes using \eqref{eq:lin}, and hence the identity takes the same form as in EFT.
Using these results one can explicitly verify that \eqref{eq:vartop} vanishes for vector and tensor gauge transformations.

\paragraph{The potential:} The scalar potential of XFT takes the following form: 
\begin{equation}
\label{eq:potXFT}
V_{\text{\tiny{XFT}}}(\mathcal{M},g,X) \,=\, V_{\text{\tiny{EFT}}}(\mathcal{M},g) \,+\, V_{\text{\tiny{SUGRA}}}(\mathcal{M},X) \,+\, V_{\text{\tiny{cross}}}(\mathcal{M},X) \,,
\end{equation}
where the scalar potential of EFT is independent of the $X$ deformation
\begin{formula}
\label{eq:potEFT}
V_{\text{\tiny{EFT}}} =&\,-\frac{1}{48}\,\mathcal{M}^{MN}\,\partial_{M}\mathcal{M}^{KL}\,\partial_{N}\mathcal{M}_{KL}+\frac12\,\mathcal{M}^{MN}\,\partial_M\mathcal{M}^{KL}\,\partial_{L}\mathcal{M}_{NK} \\[2mm]
&\,-\frac12 \,g^{-1}\partial_M g\,\partial_N\mathcal{M}^{MN}-\frac14\,\mathcal{M}^{MN}\,g^{-1}\partial_M g\,g^{-1}\partial_{N} g-\frac14\,\mathcal{M}^{MN}\,\partial_M g^{\mu\nu}\,\partial_N g_{\mu\nu} \,,
\end{formula}
while the parts exclusive to XFT are 
\begin{align}
\label{eq:posture}
V_{\text{\tiny{SUGRA}}}=\,&\frac{1}{168} \, \big[ \, X_{MN}{}^P X_{QR}{}^S \mathcal{M}^{MQ}\mathcal{M}{}^{NR}\mathcal{M}_{PS}+7X_{MN}{}^P X_{QP}{}^N\mathcal{M}^{MQ} \, \big] \, ,
\end{align}
and
\begin{align}
\label{eq:potCross}
V_{\text{\tiny{cross}}}=&\,\frac{1}{12} \, \mathcal{M}^{MN}\mathcal{M}^{KL} X_{MK}{}^P\,\partial_N \mathcal{M}_{PL} \,.
\end{align}
The full potential boils down to the one of EFT when the $X$ deformation is set to zero. Additionally, it precisely reduces to the potential of gauged maximal supergravity \eqref{eq:posture} when the fields are taken to be $y^M$-independent.%
\footnote{The different normalisation of $\,V_{\text{\tiny{SUGRA}}}\,$ with respect to ref.~\cite{deWit:2007mt} is due to the different normalisation of the Einstein--Hilbert term.} 
The term in \eqref{eq:potCross} is a purely novel feature as it is absent in both EFT and gauged maximal supergravity.

We finally give a few guidelines on the construction of the XFT potential. The various terms and coefficients in \eqref{eq:potXFT} are uniquely determined by requiring invariance under vector gauge transformations up to boundary terms, while each term is manifestly invariant under tensor gauge transformations. Throughout the computation, one has to repeatedly make use of the section constraint, the linear (or representation) constraint and the $X$-constraint. The starting point is the variation of the EFT potential under vector gauge transformations which can easily be computed using \eqref{eq:varscal} and 
\begin{formula}
\delta_{\Lambda}(\partial_M\mathcal{M}_{KL}) = &\,\, \widetilde{\mathbb{L}}_{\Lambda}(\partial_M\mathcal{M}_{KL})+2\,\mathcal{M}_{N(K}\,\partial_{L)}\partial_M \Lambda^N+\mathcal{M}_{KL}\,\partial_{M}\partial_{N}\Lambda^N \\
 &\,-2\, Y^{QR}{}_{N(K}\mathcal{M}_{L)Q}\,\partial_M\partial_R\, \Lambda^N+2 \,X_{N(K}{}^Q\mathcal{M}_{L)Q}\,\partial_{M}\Lambda^N \,,
\end{formula}
where $\,\lambda(\partial_M\mathcal{M}_{KL})=\tfrac12\,$. After the cancellations described in ref.~\cite{Hohm:2013uia}, the only non-covariant variations remaining are the ones depending (linearly) on the $X$ deformation. In order to cancel them, one needs to add counterterms to the potential which are of first order in the derivatives and the $X$. The only term\footnote{Up to equivalent rewriting using the linear constraint for the $X$ deformation.} of this type which does not vanish by virtue of the various constraints is \eqref{eq:potCross}. At this stage of the computation, it is important to realise that both the $X$ and the combination $\mathcal{M}^{-1}X\mathcal{M}$ take value in the $\textrm{E}_{7(7)}$ Lie algebra. Consequently, the adjoint projector satisfies
\begin{formula}
(\mathbb{P}_{\mathbf{133}})^M{}_N{}^K{}_L\,X_{PK}{}^L =&\, X_{PN}{}^M \, , \\
(\mathbb{P}_{\mathbf{133}})^M{}_N{}^K{}_L\,\mathcal{M}^{LP}X_{QP}{}^R\mathcal{M}_{RK}  = &\,\,\mathcal{M}^{MP}X_{QP}{}^R\mathcal{M}_{RN} \, .
\end{formula}
The vector gauge transformation of \eqref{eq:potCross} also yields additional non-covariant variations which are quadratic in $X$. These must be cancelled by extending further the potential with counterterms quadratic in $X$ and that do not contain derivatives. It again turns out that \eqref{eq:posture} are the only non-vanishing terms of this type.

\section{Relation to EFT and (non-)geometry}
\label{sec:non-geometry}

The main focus of this article is to describe deformations of EFT which are able to capture the exceptional generalised geometry of massive IIA supergravity.
It is interesting to note that the Romans mass was implemented in DFT non-geometrically by introducing a dependence of a RR potential on a dual (winding) coordinate \cite{Hohm:2011cp}.
This was made possible by the observation that RR fields in DFT only need to satisfy a weaker form of the section constraint in order to guarantee consistency of the theory.
It is therefore natural to try to relate this construction to the XFT framework.
A first obstacle is that in EFT all fields are packaged in \En representations, and as a consequence it is not so straightforward to relax the section constraint on what would be the RR fields.
The very distinction between NSNS and RR sectors relies on at least a partial solution of the EFT section constraint.
We will find a solution to this problem in terms of a factorisation Ansatz for the fields and parameters of the EFT theory that resembles a generalised Scherk--Schwarz (SS) Ansatz \cite{Berman:2012uy,Aldazabal:2013mya,Lee:2014mla,Hohm:2014qga}, but allows to perform a controlled, potentially non-geometric \emph{extension} of the coordinate dependence of fields and parameters rather than a truncation.
On the one hand, a disadvantage of this approach when compared to the XFT formalism is that it requires to break the section constraint of standard EFT in order to describe massive IIA supergravity, despite the fact that the latter is well-defined and entirely geometric in its own right.
For the same reason, it also becomes unclear whether the objects that are introduced in this context are globally well-defined.
On the other hand, the mapping that we now discuss allows us to elucidate how EFT admits (locally) consistent extensions to section-violating configurations, in such a way that no ten-dimensional background has been fixed yet and no truncation of degrees of freedom occurs. 
This is in striking contrast to non-geometric Scherk--Schwarz like compactifications that aim at reproducing  lower dimensional gauged supergravities.

We will also discuss the transformation properties of generalised affine connections in both the XFT context and the non-geometric EFT setting we are about to introduce, as further evidence for the consistency of these frameworks.

\subsection{The factorisation Ansatz}
\label{subsec:fact ans}

In this section we denote fields and parameters in the standard EFT theory with bold letters and the associated internal indices by $A,\,B$, and so on.
We begin by introducing a factorisation Ansatz for the vectors and gauge parameters of EFT
\begin{formula}
\label{factorisation Ansatz}
{\bf V}^A(x,y) = V^M(x,y)\,E_M{}^A(y)\ ,
\end{formula}
for some invertible matrix $\,E_M{}^A(y) \in \En\times\bbR^+\,$.%
\footnote{The indices $M,N,\ldots$ should not be regarded as `flat' in any sense. We propose the terminology `flurved'.}
With an abuse of language we will refer to $E_M{}^A(y)$ as a frame, but we will not investigate global definiteness of the construction here.
A similar factorisation Ansatz can be straightforwardly introduced for any other covariant object in EFT.\footnote{To extend the Ansatz to fields of $\bbR^+$ weight different from $\omega$, one must decompose $E_M{}^A = U_M{}^A \rho^\ell$, where $U\in\En$, $\rho>0$ and the power $\ell$ is related to the weight of the field. See for instance \cite{Hohm:2014qga}.%
\label{SS weight footnote}%
}
Equation \eqref{factorisation Ansatz} resembles a Scherk--Schwarz Ansatz, however note that we do not yet commit to a specific $y$-dependence of $V^M(x,y)$ and $E_M{}^A(y)$.
In particular the coefficients $V^M(x,y)$ are still allowed to depend on the internal coordinates.
In any case, given the consistency conditions that we will introduce shortly, one still obtains that the frame $E_M{}^A(y)$ factorises out of any relevant EFT expression, leaving us with a theory based on the coefficient fields.
If we impose the section constraint on ${\bf V}^A(x,y)$ and other tensors, then $V^M(x,y)$ and $E_M{}^A(y)$ are restricted to depend on the same set of internal coordinates, and \eqref{factorisation Ansatz} reduces to local field and parameter redefinitions.
Instead, we will relax the constraint and impose an alternative set of conditions that guarantee (local) consistency.

A Weitzenb\"ock connection and the corresponding torsion are associated 
to the frame $`E_M^A $ as follows \cite{Aldazabal:2013mya,Cederwall:2013naa}:
\begin{formula}
	`W_AB^C = \partial_A`E_B^M `E_M^C \ ,\qquad
	T(W)`{}_AB^C = 2`W_[AB]^C + `Y^CD_EB `W_DA^E \ ,
\end{formula}
where $\,`E_M^A `E_A^N = \delta_M^N \,$ and $\,`E_A^M `E_M^B = \delta_A^B \,$.
For the purpose of this paper we require that the induced torsion with XFT indices is constant and entirely contained in the $\cR_X$ representation, so that
\begin{formula}
\label{torsion of Weitzenbock is X}
	{T(W)}`{}_MN^P \equiv `X_MN^P \  \in\  \cR_X \ .
\end{formula}
Hence we have the identity
\begin{formula}
\label{frame Leibnitz algebra}
	\bbL_{E_M} `E_N^A = -`X_MN^P `E_P^A \ ,
\end{formula}
where the XFT indices are treated as spectators by the Lie derivative.
This indicates that the vectors $`E_M^A $ give rise to a Leibnitz algebra under the EFT Lie derivative.

In order to make contact with the construction of the previous sections we need to impose a constraint on the coefficient $V^M(x,y)$ in the factorisation Ansatz.
In fact, XFT contains partial derivatives $\partial_M$ which have not appeared in EFT yet.
We thus require
\begin{formula}
\label{E-constraint}
	`E_M^A \partial_A V^N = \delta_M^A \partial_A V^N \equiv \partial_M V^N \ ,
\end{formula}
and regard this constraint analogously to the section constraint.
Namely, as an algebraic equation on the set of coordinates on which $V^M(x,y)$ is allowed to depend, rather than as a differential equation.
We will refer to this requirement as the $E$-constraint.\footnote{There could be more general backgrounds that do not satisfy this constraint.
In such a situation the connection to an XFT framework seems unclear.}
Note that in a generalised (non-)geometric SS reduction this constraint is trivially satisfied as the coefficients in the SS expansion only depend on the external coordinates (see for instance the discussion in ref. \cite{Grana:2012rr} in the context of DFT).
If ${\bf V}^A(x,y)$ satisfies the section constraint, then the $E$-constraint implies the $X$-constraint on $V^M(x,y)$, with $`X_MN^P $ defined in \eqref{torsion of Weitzenbock is X}.
We stress that this is no longer guaranteed if $E_M{}^A(y)$ introduces a violation of the EFT section condition. 
From now on we assume that the $E$-constraint has been imposed unless otherwise specified.
Now consider a gauge parameter ${\bf\Lambda}^A(x,y) = \Lambda^M(x,y) E_M{}^A(y)$.
Then direct computation shows that
\begin{equation}
\label{factorisation Lie derivative}
	\bbL_{\bf\Lambda} {\bf V}^A = (\widetilde\bbL_\Lambda V^M) `E_M^A \ ,
\end{equation}
which already reveals the structure of the underlying XFT.
Closure of the EFT generalised Lie derivative is then granted if the XFT generalised Lie derivative closes, which reduces to the section- and $X$-constraints being imposed on $\Lambda^M$ and $V^M$, but not on $E_M{}^A$.
The factorisation property also follows for the Jacobiator, E-bracket and symmetric E-bracket, which are mapped to the corresponding expressions of the deformed theory.

Let us now make the connection between EFT and XFT more precise.
The factorisation Ansatz \eqref{factorisation Ansatz} introduces a local redundancy, because we can write
\begin{formula}
	V^M(x,y) &\to V'^M(x,y) = V^N(x,y)\, Q^{-1}(y)`{}_N^M \,,\\
	E(y)`{}_M^A &\to E'(y)`{}_M^A = Q(y)`{}_M^N E(y)`{}_N^A \,,\qquad\quad Q(y)`{}_M^N \in\En\times\bbR^+ \,,
\end{formula}
provided that $V'^M$ still solves the section-, $X$- and $E$-constraints, and that $E'`{}_M^A $ satisfies \eqref{frame Leibnitz algebra} for the \emph{same} $`X_MN^P $.
Denoting the associated infinitesimal transformation $q(y)`{}_M^N $, the latter requirement translates into the equation
\begin{formula}
\label{q condition}
	`q_A^D `X_DB^C +`q_A^D `X_AD^C -`X_AB^D `q_D^C 
	-2`\partial_[A `q_B]^C  + `Y^CD_EB `\partial_D `q_A^E    = 0\,,
\end{formula}
where we made use of the $E$-constraint.
We must gauge-fix these $q$-transformations in order to lift the redundancy introduced in the factorisation.
To this end we note that under a generalised diffeomorphism generated by ${\bf\Lambda}^A(x,y) = \Lambda^M(x,y) E_M{}^A(y)$ the frame transforms as
\begin{formula}
	\delta_{\bf\Lambda} `E_M^A &= \bbL_{\bf\Lambda} `E_M^A = -q[\Lambda]`{}_M^N `E_N^A \,,\\
	q[\Lambda]`{}_M^N &\equiv \partial_M \Lambda^N -`Y^NP_QM \partial_P\Lambda^Q +\Lambda^P `X_PM^N \,,
\end{formula}
and $q[\Lambda]`{}_M^N $ satisfies \eqref{q condition} together with all the coordinate constraints.
For any parameter ${\bf\Lambda}^A$ we can therefore define an improved variation
\begin{formula}
\label{improved variation}
\hat\delta_{\bf\Lambda} \equiv \delta_{\bf\Lambda} + \delta_{q[\Lambda]}\,\,\,\,\, , \qquad \Lambda^M = {\bf\Lambda}^A `E_A^M \,\,\,\, ,
\end{formula}
such that if ${\bf\Lambda}^A$ satisfies the factorisation Ansatz and the associated coordinate constraints, then $\hat\delta_{\bf\Lambda} `E_M^A = 0$.%
\footnote{This procedure is analogous to the construction of general coordinate transformations compensated by local Lorentz ones, such that under an isometry $\,\xi^\mu\,$, $\,\hat\delta_\xi e_\mu{}^a = 0\,$.}
Now, under a generalised diffeomorphism in EFT, $V^M$ transforms as a scalar: $\delta_{\bf\Lambda}V^M = {\bf\Lambda}^A \partial_A V^M$.
This implies that under the improved variation it  transforms as 
\begin{formula}
	\hat\delta_{\bf\Lambda} V^M = {\bf\Lambda}^A \partial_A V^M - V^N q[\Lambda]`{}_N^M = \widetilde\bbL_\Lambda V^M \,,
\end{formula}
which reproduces the generalised Lie derivative of the deformed theory.
We can thus reconstruct the geometry of XFT from the factorisation \eqref{factorisation Ansatz} and the improved variations \eqref{improved variation}.
In particular, we can define the XFT general diffeomorphism transformations from the EFT ones and from the $q$-transformations associated with the ambiguity in the factorisation Ansatz:
\begin{formula}
	\delta_\Lambda \equiv \hat\delta_{\bf\Lambda}\ .
\end{formula}

As a final remark, note that this construction can be used as a cross--check for the completeness of the action \eqref{eq:action}.
All terms in the EFT action except for the scalar potential are built from covariant objects, so that the mapping to XFT using the factorisation Ansatz is immediate.
Moreover, the decomposition of the EFT scalar potential can only yield terms with two internal derivatives and no $X$-deformation (matching the form of the original EFT potential), terms with one derivative and one $X$, and terms with no derivatives and two $X$.
All such possible terms have already been implemented in \eqref{eq:potXFT} and their coefficients are fixed by requiring invariance under internal generalised diffeomorphisms.

\subsection{Romans mass and T-dual backgrounds}

Let us now specialise to the case of the Romans mass deformation $\,X=\XR\,$.
As discussed in Section~\ref{subsec:SC}, every type II internal flux T-dual to the Romans mass can be implemented in XFT in terms of the same $\XR$ deformation, by choosing different solutions of the section and $X$-constraints.
Moreover, we saw that in $D=4$ there is a further `dual M-theory' solution of these constraints where the same $\XR$ deformation can be interpreted as a constant Freund--Rubin parameter.
We will now show how \eqref{factorisation Ansatz} can be used to map the deformed generalised Lie derivative with $X=\XR$ to a certain background in EFT.

Since $\,\XR_{MN}{}^P \,$ vanishes after any contraction with another copy of itself, a possible choice of frame $\,E(y)`{}_M^A \,$ that generates it under \eqref{torsion of Weitzenbock is X} is
\begin{formula}
\label{E frame for RR fluxes}
	E(y)`{}_M^A = \delta_M^A -\frac1c \, y^P \, \XR_{PM}{}^A \,,
\end{formula}
where we have temporarily suspended the distinction between EFT and XFT indices.
The constant $\,c\,$ is introduced to match the normalisation of \eqref{torsion of Weitzenbock is X}.

In the $\,D=9\,$ case, there is only one coordinate entering \eqref{E frame for RR fluxes} which is the winding coordinate $\,y^3\,$ from the point of view of the `natural' (massive) IIA frame (with physical coordinate~$\,y^{2}\,$). In the $\,D=7\,$ case, the coordinates entering \eqref{E frame for RR fluxes} are the three winding coordinates $\,\tilde{y}_{\alpha}\,$ from the massive IIA viewpoint (with physical coordinates~$\,y^{\alpha 4}\,$). In the $\,D=4\,$ case, \eqref{E frame for RR fluxes} activates not only the six winding coordinates $\,y_{m8}\,$ from the point of view of the massive IIA theory (with physical coordinates~$\,y^{m7}\,$), but further includes the seventh `dual M-theory' coordinate $\,y_{78}\,$.
In all these cases (and in any other dimensions, too) the coordinates activated by \eqref{E frame for RR fluxes} constitute themselves a solution of the section and $\XR$-constraint.  In other words, the  frame $\,E(y)`{}_M^A \,$ specified by \eqref{E frame for RR fluxes} satisfies these constraints itself, and the $E$-constraint reduces to the $X$-constraint. Importantly, there is no overlap between the coordinates entering (\ref{E frame for RR fluxes}) and the physical coordinates in the `natural' (massive) type IIA solution.

The Ansatz \eqref{E frame for RR fluxes} is not unique. 
Denoting $\,y_{*}\,$ one of the $y$-coordinates entering $\,E(y)`{}_M^A \,$ in (\ref{E frame for RR fluxes}), one can also construct a frame that depends \textit{only} on that specific $\,y_{*}\,$:
\begin{formula}
\label{E frame for RR fluxes simpler}
E(y_{*})`{}_M^A \,=\, \delta_M^A -\frac1{c'} \, y_{*} \, \XR_{y_{*} M}{}^A \ .
\end{formula}
This observation is important for the following reason. Recalling the discussion in Section~\ref{sec:SL5_fluxes}, the $\XR$ deformation can be interpreted as a RR background flux (also a FR parameter in the $D=4$ case) after applying T-duality transformations. In the resulting T-dual frames, the coordinate $\,y_{*}\,$ entering (\ref{E frame for RR fluxes simpler}) can always be chosen so that it becomes part of the physical coordinates permitted by the section and $X$-constraint. As a result,
the factorisation Ansatz \eqref{factorisation Ansatz} that maps EFT and XFT becomes purely geometric.
In other words, eq.~\eqref{factorisation Ansatz} induces nothing more than local field and parameter redefinitions in the standard EFT.
Analysing its action on the EFT scalar fields, one can see that \eqref{factorisation Ansatz} combined with \eqref{E frame for RR fluxes simpler} and a solution of the constraints \emph{compatible with} $\,y_{*}\,$, induces a redefinition of some Ramond--Ramond $p$-form potential of the associated supergravity theory by a term linear in $\,y_{*}\,$.
The latter then induces the constant flux encoded in $\XR$.
When $\XR$ is identified with the FR parameter of eleven-dimensional supergravity (which is only possible in $\,D=4\,$), the $p$-form potential acquiring a linear dependence on $\,y_{*}\,$ is the internal $A_{(6)}$.

The one exception to this situation is the `natural' massive IIA frame where $\,\XR\,$ is identified with the Romans mass: there is no potential that can be redefined to introduce the constant $\,F_{(0)}\equiv m_{\rm R}\,$.
This translates into the fact that, when one chooses the solution of the XFT constraints corresponding to massive IIA supergravity, the physical coordinates are incompatible with any of the $y$-coordinates entering (\ref{E frame for RR fluxes}). Therefore, \eqref{factorisation Ansatz} and \eqref{E frame for RR fluxes simpler} necessarily introduce a \textit{non-geometric} dependence of some internal RR potentials on a winding coordinate.\footnote{In $D=4$, alternatively, one can introduce dependence on $\,y_{78}\,$, which can be regarded as a `dual M-theory' coordinate.}
This is consistent with the picture in DFT, where a similar winding dependence is introduced for a RR potential in order to generate the Romans mass \cite{Hohm:2011cp}.
Our findings in this section, after fixing the $q$-transformations, can be interpreted as a generalisation of the non-geometric construction in DFT, appropriately covariantised under \En and under the complete set of exceptional generalised diffeomorphisms.

\subsection{Affine connections in EFT and XFT}
\label{sec:aff conn}

It is natural to ask whether the modified notion of covariance introduced in XFT and/or the (possibly non-geometric) factorisation Ansatz \eqref{factorisation Ansatz} in EFT allow for the definition of consistent affine connections and thus, a notion of internal covariant derivative.
We will provide here a positive answer both directly for XFT, and for EFT backgrounds of the form specified by \eqref{factorisation Ansatz}.
Since in both cases some modifications appear with respect to the standard transformation properties of an affine connection in exceptional generalised geometry, it is convenient to discuss the two frameworks at the same time and see, as a consistency check, that the objects in XFT also descend from the EFT embedding \eqref{factorisation Ansatz}.

We will first show that for EFT backgrounds satisfying the factorisation Ansatz \eqref{factorisation Ansatz}  and the necessary coordinate constraints discussed in Section~\ref{subsec:fact ans}, it is possible to define an affine connection.
First we introduce a covariant derivative acting on a vector in EFT
\begin{formula}
\label{int cov der EFT}
	D_A {\bf V}^B = \partial_A {\bf V}^B + `\Gamma_AC^B \mathbf V^C \,,
\end{formula}
and deduce the transformation property of $\,`\Gamma_AC^B \,$ from the required covariance
\begin{formula}
\label{affine connection trf proof}
	\delta_{\bf \Lambda}D_A \mathbf V^B = \bbL_{\bf \Lambda} D_A \mathbf V^B 
	\ \Rightarrow\ 
	`{\delta_{\bf \Lambda}\Gamma}_AC^B \mathbf V^C = \bbL_{\bf \Lambda} D_A \mathbf V^B - D_A \bbL_{\bf \Lambda} \mathbf V^B \,.
\end{formula}
This is just the standard procedure to deduce the transformation of the affine connection.
The crucial requirement for \eqref{affine connection trf proof} to be consistent is that the right hand side must not contain derivatives of $\,\mathbf V^A\,$.
This is usually guaranteed by the section constraint, but also holds for the backgrounds under investigation.
Indeed, assuming that both $\,\mathbf \Lambda^A\,$ and $\,\mathbf V^A\,$ satisfy \eqref{factorisation Ansatz}, and making use of the $X$- and $E$-constraints on $\,V^M$, we deduce
\begin{formula}
\label{affine conn trf EFT}
	`{\delta_{\bf\Lambda}\Gamma}_AB^C = \hat\delta_{\bf\Lambda}\Gamma`{}_AB^C =\ 
	&\bbL_{\bf\Lambda} `\Gamma_AB^C + \partial_A\partial_B{\bf\Lambda}^C - `Y^CE_FB \partial_E\partial_A {\bf\Lambda}^F\\[.2ex]
	&+`Y^DE_FA \partial_E\mathbf \Lambda^F `W_DB^C \,.
\end{formula}
The last term is new and specific to the frame $\,E_M{}^A(y)\,$ in \eqref{factorisation Ansatz}.
If $\,{\bf\Lambda}^A\,$ and $\,\mathbf V^A\,$ satisfy the section constraint, then this extra term vanishes identically and the standard transformation rule for the generalised affine connection holds.
Note that the standard and improved ${\bf\Lambda}$-variations are equal for $`\Gamma_AB^C $.

The covariance of $\,D_A\mathbf V^B\,$ is not enough to guarantee its consistency when the section constraint is violated by $\,\mathbf V^B\,$.
Closure of the EFT generalised Lie derivative is guaranteed in this setting only if $\,D_A\mathbf V^B\,$ can be factorised similarly to \eqref{factorisation Ansatz}: 
\begin{formula}
\label{cov der XFT}
	`E_M^A D_A \mathbf V^B `E_B^N \equiv \cD_M V^N \,,
\end{formula}
and $\,\cD_M V^N$ satisfies the section, $X$- and $E$-constraints.
In this case we can further define
\begin{formula}
\label{aff conn XFT}
	\cD_M V^N = \partial_M V^N + \widetilde\Gamma`{}_MP^N V^P  \,,\qquad
	\widetilde\Gamma`{}_MN^P = `\Gamma_MN^P - `W_MN^P \,,
\end{formula}
where in the last expression EFT and XFT indices are exchanged by contraction with $\,`E_A^M \,$ and its inverse.

The right hand side of eq.~\eqref{cov der XFT} corresponds to a covariant derivative in XFT.
Let us then discuss the introduction of affine connections directly in the deformed theory.
The procedure is analogous to what we have discussed so far, but now fields and parameters directly satisfy the section and $X$-constraint.
The transformation property of $\,\widetilde\Gamma`{}_AB^C \,$ is found to be
\begin{formula}
\label{aff conn trf XFT}
\delta_\Lambda\widetilde\Gamma`{}_MN^P =
\widetilde\bbL_\Lambda \widetilde\Gamma`{}_MN^P +
\partial_M\partial_N\Lambda^P -`Y^PQ_RN \partial_Q\partial_M\Lambda^R +
\partial_M\Lambda^Q `X_QN^P \,.
\end{formula}
Note that this expression contains extra $X$-dependent terms with respect to the transformation of an affine connection in EFT in a geometric setting.
This fact reflects the different notion of covariance of XFT,  defined in terms of $\,\widetilde\bbL\,$ rather than $\,\bbL\,$.
Eq.~\eqref{aff conn trf XFT} can also be deduced from \eqref{affine conn trf EFT} by making use of the $E$-constraint, showing that these definitions are mutually consistent and that the EFT factorisation Ansatz reproduces the correct structures naturally defined \mbox{in XFT}.

We now make another observation: the torsion associated with $\,\widetilde\Gamma`{}_MN^P \,$ decomposes as
\begin{formula}
	T(\widetilde\Gamma)`{}_MN^P = 2\widetilde\Gamma`{}_[MN]^P +`Y^PQ_RN \widetilde\Gamma`{}_QM^R =
	-`X_MN^P +`E_M^A `E_N^B T(\Gamma)`{}_AB^C `E_C^P \,.
\end{formula}
This means that, given a torsionless $\,`\Gamma_MN^P \,$, we can write the XFT generalised Lie derivative also as a covariant Lie derivative:
$\,	\widetilde\bbL_\Lambda = \bbL^{\cD}_\Lambda\,$.
Finally, using the  transformation \eqref{aff conn trf XFT}, it is possible to deduce that $\,\delta_\Lambda `X_MN^P = 0\,$. This is compatible with the general construction of XFT and with \eqref{deltaX}.
In the EFT embedding, the same fact descends directly from $\,\delta_\Lambda `E_M^A \equiv \hat\delta_{\bf\Lambda} `E_M^A = 0\,$.

\section{Applications and future directions}
\label{sec:Discussion}

We close the paper with a discussion on some potential applications of the $\,\En\,$ XFT framework. An immediate one is the investigation of consistent reductions of massive IIA on non-trivial geometries. Amongst these, the study of $\,\textrm{S}^{n-1}\,$ sphere reductions to gauged maximal supergravities in $\,D=11-n\,$ dimensions is of special interest. The $\,n=7\,$ case has recently been shown to determine a consistent truncation
in refs~\cite{Guarino:2015qaa,Guarino:2015vca} where a central role was played by the duality hierarchy in the gauged maximal $\,D=4\,$ supergravity \cite{deWit:2007mt,deWit:2008ta,deWit:2008gc,Bergshoeff:2009ph}. In contrast, the $\,n=4,5\,$ cases were presented in ref.~\cite{Cvetic:2000ah} only for the massless IIA theory. Therefore, it would be very interesting to perform a systematic analysis of massive~IIA reductions on $\,\textrm{S}^{n-1}\,$ in the context of generalised Scherk--Schwarz reductions of $\,\En\,$ XFT along the lines of refs~\cite{Hohm:2014qga,Baguet:2015sma,Malek:2015hma}.

A first step in this direction is to ask under what circumstances a consistent generalised SS Ansatz for massless~IIA is automatically (\textit{i.e.} without making any modification to the Ansatz itself) a consistent Ansatz for massive~IIA.
This has been shown to be true for instance for the $\,\textrm{S}^6\,$ case \cite{Guarino:2015qaa,Guarino:2015vca}.
Generalised SS reductions of massless IIA are based on a truncation Ansatz of the form $\,V^M(x,y) = v^{\ul N}(x)\,S(y)_{\ul N}{}^M\,$ for any covariant object, with $\,S(y)_{\ul N}{}^M\in\En\times\bbR^+\,$ and the $y$-dependence being restricted to IIA coordinates.%
\footnote{Tensors of weight $\,\lambda\neq\omega\,$ are described as in footnote~\ref{SS weight footnote}.}
The frame $\,S(y)_{\ul N}{}^M\,$ must satisfy the analogue of \eqref{frame Leibnitz algebra} for some embedding tensor $\,X^{\text{\tiny (S)}}_{MN}{}^P\,$.
Generic sphere Ans\"atze of this type have been constructed in \cite{Hohm:2014qga}, so it would be useful to know when they can be implemented directly also in XFT.
In the XFT framework describing massive IIA, $\,S(y)_{\ul N}{}^M\,$ must satisfy analogous conditions,  now containing extra terms related to the $\XR$-deformation of the generalised Lie derivative, which encodes the Romans mass.
It is straightforward to see that, if we want to keep the \emph{same} $\,S(y)_{\ul N}{}^M\,$ as in the massless Ansatz, then consistency is only obtained if  $\,S(y)_{\ul N}{}^M\,$ stabilises $\,\XR_{MN}{}^P\,$ (up to a global $\,\En\times\bbR^+\,$ transformation that can be always reabsorbed).
The resulting $D$-dimensional gauged supergravity will then be based on an embedding tensor  $\,X=X^{\text{\tiny (S)}}+\XR\,$.
Truncations on $\,\textrm{S}^{n-1}\,$ down to $D=11-n$ dimensions are based on twist matrices valued in an $\,\SL(n)\times \bbR^+\,$ subgroup of $\,\En\times\bbR^+\,$.
One can check that the stabiliser of $\,\XR\,$ in $\,\En\,$ contains only an $\,\textrm{SL}(n-1)\,$ group for $\,n<7\,$, which means that the massless IIA truncation Ans\"atze on spheres of dimension lower than six are not consistent for the massive theory.%
\footnote{An alternative road is to investigate whether the deformation $\,X=X^{\text{\tiny (S)}}+{\XR}'\,$, where ${\XR}'\,$ represents a generic element in the $\,\En\,$ orbit of the $\XR$ deformation,  satisfies the quadratic constraint in (\ref{quadratic constraint}). For instance, let us focus once more on the $\,\textrm{SL}(5)\,$ XFT and its counterpart, the gauged maximal $\,D=7\,$ supergravity. In ref.~\cite{Samtleben:2005bp}, the reduction of massless IIA on $\,\textrm{S}^{3}\cong \tfrac{\textrm{SO}(4)}{\textrm{SO}(3)}\,$ was connected to an ISO(4)-gauged maximal supergravity given (in our conventions of Section~\ref{sec:SL5_fluxes}) by a deformation of the form $\,Y_{MN}=R^{-1} \, \textrm{diag$(1,1,1,0,1)$}\,$, where $\,R\,$ relates to the $\,\textrm{S}^{3}\,$ radius. Such a $\,Y_{MN}\,$ determines $\,X^{\text{\tiny (S)}}\,$. Adding now a generic element $\,{\XR}'\,$ in the SL(5) orbit of the $\,\XR\,$ deformation and computing the resulting quadratic constraints (\ref{quadratic constraint}) for $\,X=X^{\text{\tiny (S)}}+{\XR}'\,$, one finds that a consistent truncation on $\,\textrm{S}^{3}\cong \tfrac{\textrm{SO}(4)}{\textrm{SO}(3)}\,$ (finite $\,R\,$) to a maximal $\,D=7\,$ supergravity is possible only in the massless case ($m_{\rm R}=0$). Let us emphasise again that we are assuming the same reduction Ansatz both in the massive and massless cases.}
Only in $\,D=4\,$ does $\,\XR\,$ break \Eseven to $\,\SL(7)\,$ (plus a solvable piece), which shows that the massless IIA Ansatz for $\,\textrm{S}^6\,$ can be directly utilised on the massive theory, as was indeed done in \cite{Guarino:2015qaa,Guarino:2015vca}.

Also in the context of massive IIA, it would of course be interesting to implement in our formalism a way to reproduce the equations of motion and Bianchi identities associated with \emph{general} type IIA backgrounds, where the value of the Romans mass $\,m_{\rm R}\,$ is allowed to change by discrete values when crossing a D8-brane.
In order to achieve this, we proceed by allowing for a non-constant rescaling of the Romans $\,X^{\raisebox{.5ex}{\tiny R}}\,$ deformation
\begin{formula}
X^{\raisebox{.5ex}{\tiny R}}_{MN}{}^P \to M(x,y)\, X^{\raisebox{.5ex}{\tiny R}}_{MN}{}^P \ ,
\end{formula}
where $\,M(x,y)\,$ plays the role of a spacetime-dependent Romans mass.\footnote{We can set to unit value the constant $m_{\rm R}$ parameter contained in $\XR$.}
Constancy of the latter can be imposed by adding $p$-form Lagrange multipliers to the XFT action that enforce $\,{\del_\mu M = \del_M M = 0}\,$.
This approach would be equivalent to the original construction of ref.~\cite{Bergshoeff:1996ui}, if the extended coordinates $\,y^M\,$ are restricted to the massive IIA coordinates.
Such a construction can be made more general by making the full $X$ deformation $\,x\,$ and $\,y\,$ dependent and introducing Lagrange multipliers to enforce its constancy as well as its (linear and quadratic) constraints.
This is the standard approach to derive the complete tensor hierarchy of gauged supergravities \cite{Nicolai:2000sc,deWit:2008gc}, and it would be interesting to investigate the consistency of such an approach in XFT where the section constraint and the tensor hierarchy must be taken into account appropriately. The study of D8-branes in XFT, especially the interpretation of the \mbox{$X$-constraint} in (\ref{X condition}) as a projector into specific U-duality brane charges, might help in understanding mutually 1/2-BPS configurations \cite{Bossard:2015foa} in the massive IIA theory.

Moving now to the context of the type IIB theory, we saw in Section~\ref{sec:SL5_fluxes} that, together with ordinary \mbox{$p$-form} fluxes, all the $\,\En\,$ XFT's are compatible with an SL(2)-triplet of one-form deformations $\,\mathcal{H}_{A}\equiv(F,H,\hat{F})\,$.
These are connected to the triplet of eight-form potentials in type~IIB \cite{Dall'Agata:1998va}, thus becoming relevant in the study of S-duality orbits of \mbox{7-branes} \cite{Meessen:1998qm} and potentially of \mbox{F-theory}. 
An SL(2) invariant constraint guarantees that the three eight-form potentials are dual to the two scalar degrees of freedom of the IIB axion-dilaton.
The XFT consistency constraints do not impose this extra requirement.
Therefore it would be interesting to investigate if XFT allows to describe more general type IIB backgrounds, and clarify 
 whether $\,\mathcal{H}_{A}\,$ in XFT is entirely geometric or contains a bit of `global \mbox{non-geometry'} (see discussion on $\gamma$-deformations in \cite{Aldazabal:2010ef}).

Finally there are other interesting directions which are more tangential to the content of the present paper.
The first one is the construction of a supersymmetric version of the XFT's similar to the ones for the undeformed  EFT's \cite{Godazgar:2014nqa,Musaev:2014lna}. 
The analysis of Section~\ref{sec:aff conn} suggests that there should be no obstruction in defining the K$(\En)$ connections that are required for introducing fermions.
The second one is the formal truncation of the $\,\En\,$ XFT to a deformed $\,\textrm{O}(n-1,n-1)\,$ DFT. 
Such a deformed DFT should connect with the formalism introduced in \cite{Hohm:2011ex} to account for non-Abelian gauge couplings in the DFT formulation of the heterotic string, except that a non-trivial $\textrm{O}(n-1,n-1)$-valued deformation $\,f_{MN}{}^P\,$ should appear together with extra constraints (partially) reproducing the embedding tensor constraints in half-maximal supergravity. The structure of such DFT deformations must also be similar to the formalism introduced in \cite{Grana:2012rr} to describe dimensional reductions of DFT.
The crucial difference with respect to the construction in \cite{Grana:2012rr} is that no truncation of the coordinate dependence is required, thus resulting in a deformation of the generalised Lie derivative of the \textit{full} theory.
The last question concerns the existence of an $\,\textrm{E}_{8(8)}\,$ XFT and its relation to gauged maximal $\,D=3\,$ supergravity \cite{Nicolai:2000sc}. A difference in the E$_{8(8)}$ case is the presence of an extra covariantly constrainted vector gauge parameter required for closure of generalised diffeomorphisms \cite{Hohm:2014fxa} (for an alternative approach see also \cite{Rosabal:2014rga}). Investigating the potential implications of this new term on the $X$ deformation goes beyond the scope of this paper. We hope to come back to these and related questions in the near future.

\paragraph{Note added:} Shortly after this manuscript appeared on the arXiv, the preprint \cite{Cassani:2016ncu} appeared with a detailed construction of the exceptional generalised geometry for massive IIA supergravity. It reaches similar conclusions regarding sphere reductions of IIA supergravity, and further investigates alternative Ans\"atze for the massive theory.

%
%

\section*{Acknowledgments}

We would like to thank D. Butter, B. de Wit, O. Hohm, A. Kleinschmidt and D. Robbins for interesting conversations. The work of the authors is supported by the ERC Advanced Grant no. 246974, ``Supersymmetry: a window to non-perturbative physics''.

%
%

\small

\bibliography{references}
\bibliographystyle{utphys}

\end{document}

%% file: definitions.tex
\newcommand\blfootnote[1]{%
  \begingroup
  \renewcommand\thefootnote{}\footnote{#1}%
  \addtocounter{footnote}{-1}%
  \endgroup
}

\def\del{\partial}

\def\sst#1{{\scriptscriptstyle #1}}

\def\wtd{\widetilde}

\def\0{{\sst{(0)}}}
\def\1{{\sst{(1)}}}
\def\2{{\sst{(2)}}}
\def\3{{\sst{(3)}}}
\def\4{{\sst{(4)}}}
\def\5{{\sst{(5)}}}
\def\6{{\sst{(6)}}}
\def\7{{\sst{(7)}}}
\def\8{{\sst{(8)}}}

\newcommand{\ul}[1]{\underline{#1}}

\newcommand{\cD}{{\mathcal{D}}}

\newcommand{\cR}{{\mathcal{R}}}

\newcommand{\bbL}{{\mathbb{L}}}

\newcommand{\bbP}{{\mathbb{P}}}

\newcommand{\bbR}{{\mathbb{R}}}

\def\En{\ensuremath{\mathrm{E}_{n(n)}}\xspace}

\newcommand{\Eseven}{\ensuremath{\mathrm{E}_{7(7)}}\xspace}

\def\SL(#1){\ensuremath{\mathrm{SL}(#1)}}
\def\GL(#1){\ensuremath{\mathrm{GL}(#1)}}
\def\Sp(#1){\ensuremath{\mathrm{Sp}(#1)}}
\def\USp(#1){\ensuremath{\mathrm{USp}(#1)}}
\def\SU(#1){\ensuremath{\mathrm{SU}(#1)}}
\def\U(#1){\ensuremath{\mathrm{U}(#1)}}
\def\SO(#1){\ensuremath{\mathrm{SO}(#1)}}
\def\O(#1){\ensuremath{\mathrm{O}(#1)}}
\def\ISO(#1){\ensuremath{\mathrm{ISO}(#1)}}

\def\fsl(#1){\ensuremath{\mathfrak{sl}(#1)}}
\def\su(#1){\ensuremath{\mathfrak{su}(#1)}}
\def\sp(#1){\ensuremath{\mathfrak{sp}(#1)}}
\def\so(#1){\ensuremath{\mathfrak{so}(#1)}}
\def\iso(#1){\ensuremath{\mathfrak{iso}(#1)}}

\def\XR{X^{{\text{\tiny R}}\vphantom{()}}\xspace}